\documentclass[10pt, journal]{IEEEtran}

\usepackage{cite}
\usepackage{amsmath,amssymb,amsfonts}
\usepackage{graphicx}
\usepackage{textcomp}
\usepackage{subfig}
\usepackage[font=footnotesize]{caption}
\usepackage[table]{xcolor}
\usepackage{tabularx}
\usepackage{enumitem}
\usepackage{balance}
\usepackage{multirow}
\usepackage{array}
\usepackage{adjustbox}
\renewcommand{\arraystretch}{1.3}
\newcommand{\stimes}{{\times}}

\usepackage[linesnumbered, ruled, lined, hangingcomment]{algorithm2e}

\begin{document}
\bstctlcite{IEEEexample:BSTcontrol}

\title{Automated Design Approximation to Overcome Circuit Aging}

\author{
Konstantinos~Balaskas,
Georgios~Zervakis,
Hussam~Amrouch, \IEEEmembership{Member, IEEE},\\
J\"org~Henkel, \IEEEmembership{Fellow, IEEE}, 
Kostas~Siozios, \IEEEmembership{Member, IEEE}

\thanks{Manuscript received April 7, 2021; revised July 14, 2021, accepted August 16, 2021. (\textit{Corresponding author: Konstantinos Balaskas, e-mail: kompalas@physics.auth.gr}).}%
\thanks{This work is supported in parts by the German Research Foundation (DFG) project ``ACCROSS'' and by  the E.C. funded program SERRANO under H2020 Grant Agreement No: 101017168.}
\thanks{K. Balaskas and K. Siozios are with the Department of Physics, Aristotle University of Thessaloniki, Thessaloniki 54124, Greece.}%
\thanks{H.~Amrouch is with the Chair for Semiconductor Test and Reliability (STAR) at University of Stuttgart, Stuttgart 70174, Germany.}%
\thanks{G. Zervakis and J. Henkel are with the Chair for Embedded Systems at Karlsruhe Institute of Technology, Karlsruhe 76131, Germany.}%
}

\markboth{Published in IEEE Transactions on Circuits and Systems I: Regular Papers. DOI: 10.1109/TCSI.2021.3106149}%
{K. Balaskas \MakeLowercase{\textit{et al.}}: Automated Design Approximation to Suppress Circuit Aging}

\maketitle

\begin{abstract}\boldmath
Transistor aging phenomena manifest themselves as degradations in the main electrical characteristics of transistors.
Over time, they result in a significant increase of cell propagation delay, leading to errors due to timing violations, since the operating frequency becomes unsustainable as the circuit ages. 
Conventional techniques employ timing guardbands to mitigate aging-induced delay increase, which leads to considerable  performance losses from the beginning of the circuit's lifetime.
Leveraging the inherent error resilience of a vast number of application domains, approximate computing was recently introduced as an aging mitigation mechanism.
In this work, we present the first automated framework for generating \textit{aging-aware approximate circuits}.
Our framework, by applying directed gate-level netlist approximation, induces a small functional error and recovers the delay degradation due to aging.
As a result, our optimized circuits eliminate aging-induced timing errors.
Experimental evaluation over a variety of arithmetic circuits and image processing benchmarks demonstrates that for an average error of merely $5\times10^{-3}$, our framework completely eliminates aging-induced timing guardbands.
Compared to the respective baseline circuits without timing guardbands (i.e.,~iso-performance evaluation), the error of the circuits generated by our framework is $1208$x smaller.
\end{abstract}


\section{Introduction}\label{sec:intro}

\IEEEPARstart{E}{xploiting} the inherent error resilience of a vast number of application domains, \textit{approximate computing} is established as a promising design paradigm to boost the efficiency of our computing systems~\cite{Han:ETS:2013}.
Such applications, e.g., image and digital signal processing~\cite{Han:ETS:2013,Pashaeifar:TCASI:2018,Soares:TCASI:2019,Shafique:DAC2015,Strollo:TCASI:2020,Bhardwaj:ISQED:2014,Zervakis:TVLSI:2016,Mrazek:DATE:2017,Schlachter:TransVLSI:2017,Zervakis:TVLSI:2019:vader,lingamneni:TECS:2013,Zervakis:IEEEACC:2020,Jain:DATE:2016,Nepal:DATE:2014,Venkataramani:TCADICS:2020:salsa2,Sayandip:ISLPED:2019,Kim:TCASI:2020,Amrouch:DAC:2017,Amrouch:DAC:2016:aglib,Salamin:DATE:2021,Bouroujerdian:ICCD:2018,Chen:TMSCS:2017,Paim:TCSVT:2019,Paim:TCASI:2021} and neural networks~\cite{ Tasoulas:TCASI:2020,Mrazek:ICCAD:2019}, are able to produce results of acceptable quality, while tolerating errors in the underlying computations.
Concisely, approximate computing intelligently trades off implementation/computational accuracy for gains in other metrics (e.g., energy consumption)~\cite{Han:ETS:2013}.

Leveraging this potential for efficiency improvement, approximate circuits design has gained significant research interest.
Arithmetic circuits, like adders~\cite{Pashaeifar:TCASI:2018,Soares:TCASI:2019,Shafique:DAC2015} and multipliers~\cite{Strollo:TCASI:2020,Zervakis:TVLSI:2016,Bhardwaj:ISQED:2014}, are mainly targeted as they constitute the fundamental building blocks of DSP and error resilient applications~\cite{Zervakis:TVLSI:2019:vader}.
In addition, complex approximate hardware accelerators are also proposed, such as coordinate rotation digital computer (CORDIC)~\cite{Chen:TMSCS:2017}, H.265/HEVC encoder~\cite{Paim:TCSVT:2019,Paim:TCASI:2021} and Deep Neural Networks~\cite{Mrazek:ICCAD:2019,Tasoulas:TCASI:2020}.
However, such approaches are application specific, limiting their applicability.
It is noteworthy that approximate computing exacerbates the design complexity by introducing a new dimension in systems design, i.e., the error~\cite{Han:ETS:2013}. 
To mitigate the increased complexity of approximate design and enable circuit-independent approximations, several automated methodologies are presented~\cite{Zervakis:IEEEACC:2020,Zervakis:TVLSI:2019:vader,Schlachter:TransVLSI:2017,lingamneni:TECS:2013}.

Despite the significant energy benefits derived by approximate design, recent research works~\cite{Kim:TCASI:2020,Amrouch:DAC:2017,Amrouch:DAC:2016:aglib,Salamin:DATE:2021} have gone a step forward and demonstrated that approximate computing can be also employed to efficiently suppress circuit aging effects.
Transistor aging phenomena, e.g., Bias Temperature Instability (BTI) and Hot-Carrier Injection (HCI), manifest themselves as degradations in the main electrical characteristics of transistors.
The most important degradation is related to increased threshold voltage ($V_{th}$)~\cite{Mahapatra:TED:2013}, which in turn leads to a decrease in the drain current of the transistor in the ON state ($I_{ON}$).
The latter increases significantly the transistor's propagation delay.
Therefore, the standard cells become slower over time~\cite{hu:2010}. 
Hence, circuits start to exhibit timing errors because the operating frequency becomes unsustainable as the circuit ages.
Aging-induced timing errors, are by nature very large and catastrophic for the application's accuracy since they mainly affect the most significant bits~\cite{Amrouch:DAC:2016:aglib,Salamin:DATE:2021}.
In addition, they are unpredictable and hard to control since they depend on the sequence of inputs and the state of the system~\cite{Salamin:DATE:2021,Zervakis:TVLSI:2019:vader}.

In order to keep aging effects at bay for the entire projected lifetime (e.g., 10 years), a timing guardband ($t_{GB}$) is included on top of the critical path delay at design time.
As a result, timing violations are prevented, but performance is degraded, as the circuit is forced to operate at a lower frequency even at the early phases of its lifetime, when aging-induced degradations are still negligible.
Timing guardbands are calculated at a worst case scenario, depriving the opportunity to operate the circuit at a higher frequency during the earlier part of its life cycle~\cite{Amrouch:DAC:2017}.
Several approaches have been proposed~\cite{Roy:CADICS:2016,Keane:Spectr:2011} regarding $t_{GB}$ narrowing and aimed to minimize the associated performance losses.
On the other hand,~\cite{Kim:TCASI:2020,Amrouch:DAC:2017,Amrouch:DAC:2016:aglib,Bouroujerdian:ICCD:2018,Salamin:DATE:2021} apply approximate computing to reduce the circuit's critical path delay such that operation at the maximum frequency (of the fresh exact circuit) is guaranteed throughout the circuit's lifetime.
Hence, aging-induced timing errors are eliminated by inducing smaller, deterministic, and predictable functional errors, reestablishing, thus, the reliability of the circuit \cite{Kim:TCASI:2020}.
However, despite their promising results,~\cite{Kim:TCASI:2020,Amrouch:DAC:2017,Amrouch:DAC:2016:aglib,Bouroujerdian:ICCD:2018,Salamin:DATE:2021} target very simple topologies such as ripple-carry adders and array multipliers, limiting their exploitation and providing only some preliminary, non-conclusive results regarding the exploitation of approximate computing to suppress circuit aging.

In this work, we propose for the first time an automated, circuit-agnostic design framework that generates aging-aware approximate combinational circuits.
Our proposed and implemented framework employs a genetic algorithm to concurrently apply, in a coordinated manner, wire-by-wire~\cite{Venkataramani:DATE:2013} and wire-by-constant~\cite{Schlachter:TransVLSI:2017} replacements, managing to completely eliminate the aging induced timing guardband while minimizing the induced logic error.
Unlike state-of-the-art aging-aware combinatorial circuit design automation techniques, our framework enables applying aging-aware approximation and by introducing a static error at design time, it efficiently eliminates timing errors due to aging.
Still, state-of-the-art techniques for aging-aware combinatorial circuit design are orthogonal to our work and can be applied synergistically.
Our framework operates on the circuit's post-synthesis gate-level netlist and thus it is circuit-agnostic and can be seamlessly integrated into any standard design flow.
In addition, our framework provides a scalable and time-efficient solution since it employs high-level delay and error models and thus it fully exploits the inherent parallelism of genetic algorithms.
Unlike the state-of-the-art in aging-driven approximation that examines only very simple topologies and applies handcrafted solutions, we evaluate our framework over four speed-optimized adders and multipliers from the industry-level Synopsys DesignWare library as well as five image processing benchmarks implemented with DesignWare components.
Our evaluation demonstrates that by inducing a negligible functional error (in terms of Normalized Mean Error Distance (NMED)), aging guardbands can be removed for the entire projected lifetime.

\textbf{Our novel contributions within this paper are as follows:}
\begin{enumerate}[wide, labelwidth=!, labelindent=0pt, label=\textbf{(\arabic*)}, ref=(\arabic*)]
    \item
    We present the first automated design framework for applying aging-aware approximation on combinational circuits.
    \item
    This is the first work that applies wire-by-wire and wire-by-constant approximation concurrently and in a coordinated manner and enables delay-optimized netlist approximations.
    \item
    We demonstrate that for an average error (NMED) of merely $5\times 10^{-3}$, our framework eliminates the performance loss due to aging guardbands, while aging-induced timing errors are suppressed for the entire lifetime of 10 years.
    \item
     We demonstrate that for the projected lifetime, compared to the baseline circuits without timing guardbands, the error (NMED) of the approximate circuits generated by our framework is $1208$x smaller.
\end{enumerate}

\section{Related Work}\label{sec:rel}

Targeting to alleviate the increased complexity introduced by approximate hardware design, many approximate computing works focus on the automation of the approximate circuit generation.
In \cite{Venkataramani:TCADICS:2020:salsa2}, a methodology to create approximate designs by identifying the approximation ``don't cares" within a quality constraint circuit and using them for logic simplification is proposed.
The authors in \cite{Venkataramani:DATE:2013} utilize similarities between nodes in a circuit implementation and introduce a substitution technique to produce approximate and quality configurable designs in an automated manner.
Quality configurable circuits are also generated in \cite{Jain:DATE:2016}, through an approximation design framework that identifies specific logic islands to be isolated at runtime, depending on a given quality constraint.
Netlist modification with approximate standard cells is presented in \cite{Sayandip:ISLPED:2019}, where a heuristic model is employed for the generation of approximate arithmetic circuits in reasonable design time.
In~\cite{Mrazek:DATE:2017} and~\cite{Ceska:ICCAD:2017}, cartesian genetic programming is used to generate approximate multipliers and adders in an automated manner.
The authors in~\cite{Leipnitz:TECS:2019} extend high-level synthesis (HLS) tools and apply variable-to-constant approximation to generate approximate circuits that satisfy real-time constraints.
The authors in~\cite{Schlachter:TransVLSI:2017} proposed an automated framework for generating approximate arithmetic circuits that achieve considerable energy savings.
In~\cite{Schlachter:TransVLSI:2017}, gate-level pruning of the post-synthesis netlist is applied by removing netlist's gates based on their significance and activity.
The works proposed in~\cite{Zervakis:IEEEACC:2020} and~\cite{Zervakis:TVLSI:2019:vader} extend~\cite{Schlachter:TransVLSI:2017} and enable runtime reconfigurability~\cite{Zervakis:IEEEACC:2020} or boost energy gains~\cite{Zervakis:TVLSI:2019:vader}.
In~\cite{Zervakis:IEEEACC:2020} the approximated gates are not pruned but replaced by switches, while in~\cite{Zervakis:TVLSI:2019:vader} a voltage-driven netlist pruning scheme is proposed, aiming to minimize the voltage supply value.
In~\cite{Castro:ICCAD:2020:DSEwam}, a library of approximate functional units combined with analytical models are used to build an automated framework for HLS of approximate accelerators.
All the aforementioned works target mainly power and not delay optimization and do not consider aging-specific approximations.

The authors in~\cite{Ebrahimi:ICCAD:2013} introduced a technique to optimize the circuit against aging by capturing  potential paths that may become critical after aging, and then applied iteratively tighter timing constraints on them to force the synthesis tool to optimize these paths.  
Nevertheless, the structure of paths may change and thus different gates will be used that might be more susceptible to aging.
~\cite{Kiamehr:DATE:2014} proposed to redesign the library celss through adjusting the $W_{PMOS}/W_{NMOS}$ ratio to keep the rise and fall delays balanced.
However, one cannot expect a unique sizing for each gate/cell.
Moreover, to mitigate aging effects, there are many techniques (e.g.,~\cite{Gunadi:MICRO:2010}) that aim to balance the duty cycle.
In general, such techniques are orthogonal to our work. 
When complex designs like processors are targeted, aging-aware circuit design techniques often analyze the impact of aging on the critical path only, e.g.,~\cite{Karimi:TACO:2015}.
However, aging may switch a path from critical to uncritical and vice versa, as it has been demonstrated in~\cite{Amrouch:DAC:2016:aglib}.
Other works, e.g.,~\cite{Gnad:DAC:2015}, proposed to consider the top $x$\% of the critical path.  
This might not be feasible in realistic designs since the number of paths within the top $5$\% may reach $>$~$10^{7}$~\cite{Ebrahimi:ICCAD:2013}.
In practice, determining an $x$ such that it is guaranteed that the path that may become critical after aging is included is not trivial.

The authors in~\cite{Amrouch:DAC:2016:aglib} proposed the integration of degradation-aware cell libraries in the design flow to effectively mitigate aging effects and enable  fine-grained timing guardbands.
In~\cite{Amrouch:DAC:2017}, the authors introduce the concept of aging-aware approximation.
\cite{Amrouch:DAC:2017} applies precision scaling on the inputs of a ripple carry adder, demonstrating that approximation can mitigate the aging-induced timing errors.
\cite{Amrouch:DAC:2017} demonstrated that an approximate-aware circuit to overcome aging provides around $11$\% - $13$\% less area, energy and delay overheads compared to state-of-the-art aging-aware combinatorial circuit design technique.
Despite encouraging results even for 10 years of aging, their methodology lacks the guarantee of eliminating timing guardbands.
Extending~\cite{Amrouch:DAC:2017}, a dynamic approach to substitute timing errors caused by aging with computation approximation errors was presented in~\cite{Kim:TCASI:2020}.
A detailed monitoring scheme of the critical path allowed for dynamic adaptation of the inserted approximations whenever timing violations would be detected.
The methodology however is circuit-specific and also introduces small guardbands for mismatch correction between the monitoring system and the actual circuit.
Both~\cite{Amrouch:DAC:2017} and~\cite{Kim:TCASI:2020} are applied on very simple topologies, i.e., the slow ripple carry adder and array multiplier, limiting their exploitation.
\cite{Sato:IEICE:2020} employs reconfigurable approximate circuits to compensate for aging-induced timing violations. By monitoring the performance degradation of a subject circuit throughout its lifetime with an aging sensor (i.e. a Canary FF in their implementation), the authors propose switching from accurate to approximate operation mode when timing violations are detected.
The efficiency of~\cite{Sato:IEICE:2020} heavily relies on the reconfigurability of the employed approximate circuits and the respective trade-offs they offer in terms of accuracy and performance.
Similar to~\cite{Amrouch:DAC:2017,Kim:TCASI:2020}, the approach of~\cite{Sato:IEICE:2020} is evaluated only on a simple adder circuit.
In \cite{Bouroujerdian:ICCD:2018}, a dynamic approach is proposed that trades off pessimistic thermal timing guardbands for quality and delay configurable approximate circuits.
The runtime reconfigurable designs extend over two approximation levels and introduce notable speedups with minimal quality loss.
However, significant overheads in terms of area are introduced.
Finally, \cite{Salamin:DATE:2021} applies dynamic approximation through input compression to eliminate timing guardbands.
Nevertheless, the approach presented in~\cite{Salamin:DATE:2021} is specific for DNN accelerators.

Our work distinguishes from existing state of the art as follows:
This is the first automated framework that applies aging-aware approximation.
Our framework is circuit-agnostic and can seamlessly extend any typical design flow by operating on the synthesized gate-level netlist.
Our framework completely eliminates aging-induced timing guardbands and doesn't add any area or power overheads.

\section{Aging-Aware Cell Library Characterization}\label{sec:aginglibs}
In order to estimate how transistor aging ultimately impacts the delay of a circuit's path, we employ aging-aware standard cell libraries.
Those libraries are characterized so that every transistor exhibits the worst-case aging-induced degradation (i.e.,~threshold voltage increase).
This provides us with maximum delay increase that circuits' paths might exhibit under aging for the entire projected lifetime.
As mentioned earlier, transistor aging phenomena such as BTI and HCI, induce different types of defects during the lifetime.
Those defects accumulate at the interfacial layer Si-SiO$_2$ and within the transistor dielectric. 
Over time, they weaken the electric field and hence result in an increase in the transistor threshold voltage ($\Delta V_{th}$). 
To estimate the latter, we employ a physics-based aging model validated against measurements~\cite{amrouch_TED18, sami_TVLSI19}. 
To capture wort-case aging, we consider a maximum shift in the transistor threshold voltage of $50$mV, which is considered a critical degradation in many industrial applications~\cite{BTI_book}. 
In this work, we consider the $45$nm technology node and we employ the open-source $45$nm standard cell library~\cite{nangate}, along with the Predictive Technology Model (PTM)~\cite{ptm, ptm_web} to model the underlying pMOS and nMOS transistors.
The industry standard compact model BSIM6~\cite{bsim6} for planer transistors was used in Synopsys HSPICE simulations during the library characterization. 

After estimating the worst-case $\Delta V_{th}$, we modify the transistor model and then we perform library characterization using Synopsys SiliconSmart tool flows~\cite{synopsys}. To this end, we employ the SPICE netlists of standard cells including parasitics and the degraded transistor models. For every standard cell, we consider $7 \times 7$ input signal slews and output load capacitances, similar to commercial standard cell libraries. 

The generated standard cell libraries contain the manner in which all cells are impacted by aging-induced degradation.
The open source $45$nm PDK~\cite{nangate} was used for both the baseline as well as the aging-aware libraries, thus enabling direct comparisons and evaluations.
Our libraries are fully compatible with existing standard tool flows for logic synthesis and timing analysis.
Therefore, they can be directly deployed without any modification and utilized to extract how the delay of circuits' paths will be affected by transistor aging.

\begin{figure}[!t]
    \centering
    \includegraphics[width=.48\textwidth]{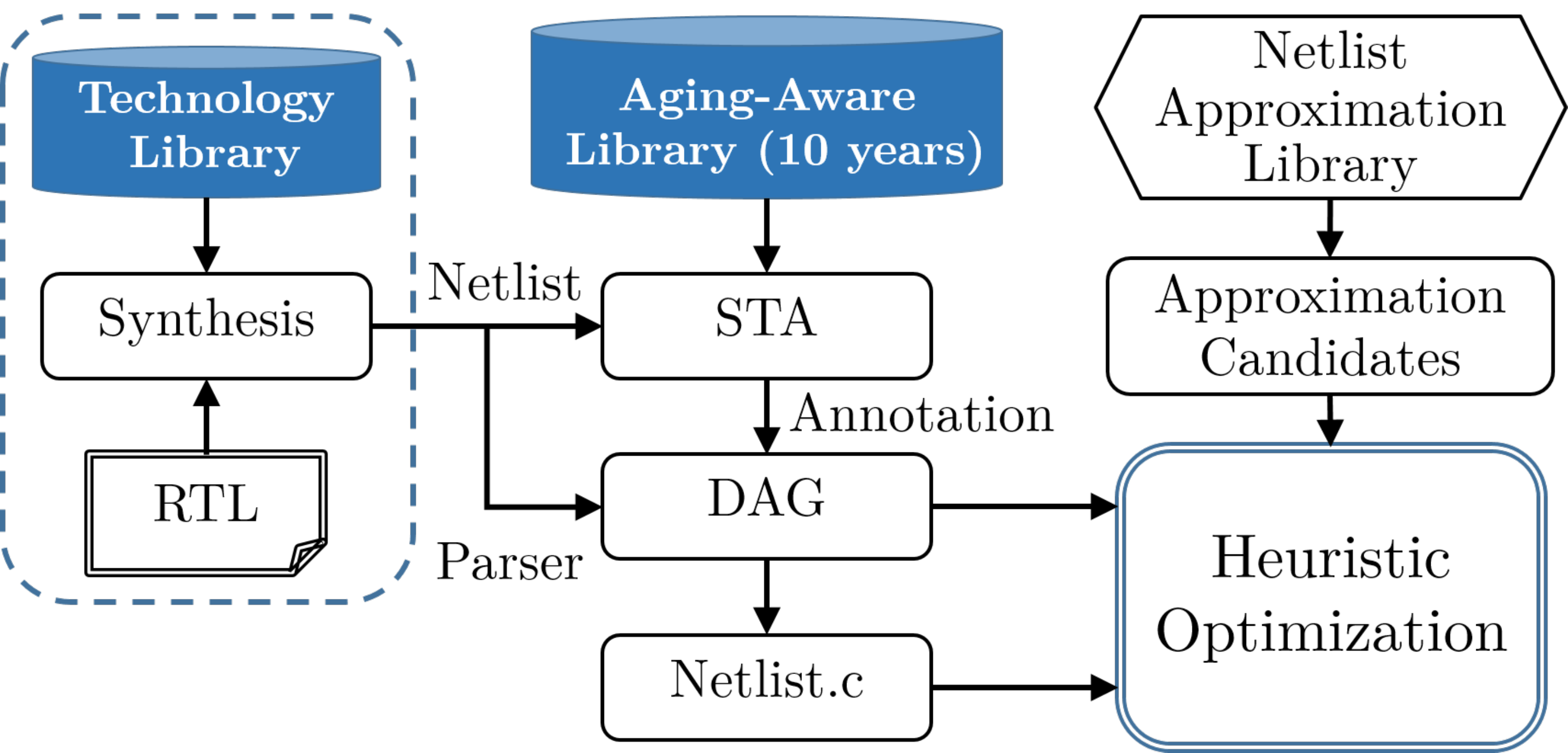}
    \caption{Abstract overview of our proposed framework that applies aging-aware approximation.}
    \label{fig:flow}
\end{figure}

\section{Proposed Framework}\label{sec:framew}
In this section, our proposed aging-aware approximation framework (illustrated in Fig.~\ref{fig:flow}), is described.
Our framework takes as input the post-synthesis gate-level netlist of the examined circuit.
Without any loss of generality, if the netlist is not available, an EDA tool is used to synthesize the Behavioural/RTL description of the circuit and generate the gate-level netlist.
The reason for operating over the post-synthesis netlist is to make our framework automated and circuit-agnostic since post-synthesis netlists feature a regular structure and our framework need only to comprise a grammar to detect the gates of the considered technology library.
This netlist will be referred hereafter as baseline netlist.
Using our aging-aware library, we perform a Static Timing Analysis (STA) to obtain the timing information of the respective aged circuit.
Next, based on the obtained aging-aware timing analysis, we run our optimization phase to generate the approximate netlist.
The goal of our optimization is to identify the respective approximations so that
i) the critical path delay (CPD) of the aged approximate netlist is less or equal to the CPD of the baseline fresh (i.e., w/o aging) netlist
and ii) the error of the approximate netlist is minimized:
\begin{equation}\label{eq:optimization}
\begin{gathered}
\text{given } fresh\, BL\, netlist \text{ find } aged\,AX\,netlist \\
\text{s.t.}\,\mathrm{CPD}(aged\,AX\,netlist) \leq \mathrm{CPD}(fresh\, BL\, netlist)\\
\text{and}\,\min\big(Error(aged\,AX\,netlist)\big)
\end{gathered}
\end{equation}
In other words, the approximate netlist can operate from day zero (fresh) until the end of the projected lifetime ($50$mV $\Delta V_{th}$ degradation\footnote{In our work, we protect the circuit against a $V_{th}$ increase of a maximum $50$mV, which is considered to be a critical level of degradation \cite{BTI_book}.} \cite{BTI_book})
at the CPD of the fresh baseline netlist with error guaranties, i.e., no aging-induced timing errors will occur.
Hence, the applied approximation completely eliminates the aging-induced timing guardbands.
In addition, the error of the approximate netlist is constant (and predictable since our framework introduces a functional/logic error) for the entire projected lifetime, i.e., the output of the approximate netlist is not affected by aging-induced degradations.
Such a degradation (i.e., $\Delta V_{th}$ = $50$mV) can be reached after $10$ years, or much earlier at around $3$ years and less, depending on the operating condition of the circuit (e.g., workload induced stress, operating temperature, voltage, etc.)~\cite{BTI_T_Vdd}.
Still, our technique is orthogonal to the specific degradation level since the only modification required is to characterize our standard cell library at the desired degradation level.
To generate the approximate netlist our framework uses a library of netlist approximation techniques and employs a genetic algorithm.
Genetic algorithms are proven to deliver close to optimal solutions in a fast manner, due to their inherent parallelism.
Hence, note that our framework applies a static approximation, at design time, and by inducing a small error, it generates approximate circuits that are by definition designed to tolerate such a $\Delta V_{th}$ degradation, as~\eqref{eq:optimization} reveals.
Finally, the obtained approximate netlist is synthesized to exploit all the optimizations performed by the synthesis algorithms.

\begin{figure}[!t]
    \centering
    \subfloat{
    \includegraphics[height=.45\textwidth, angle=90]{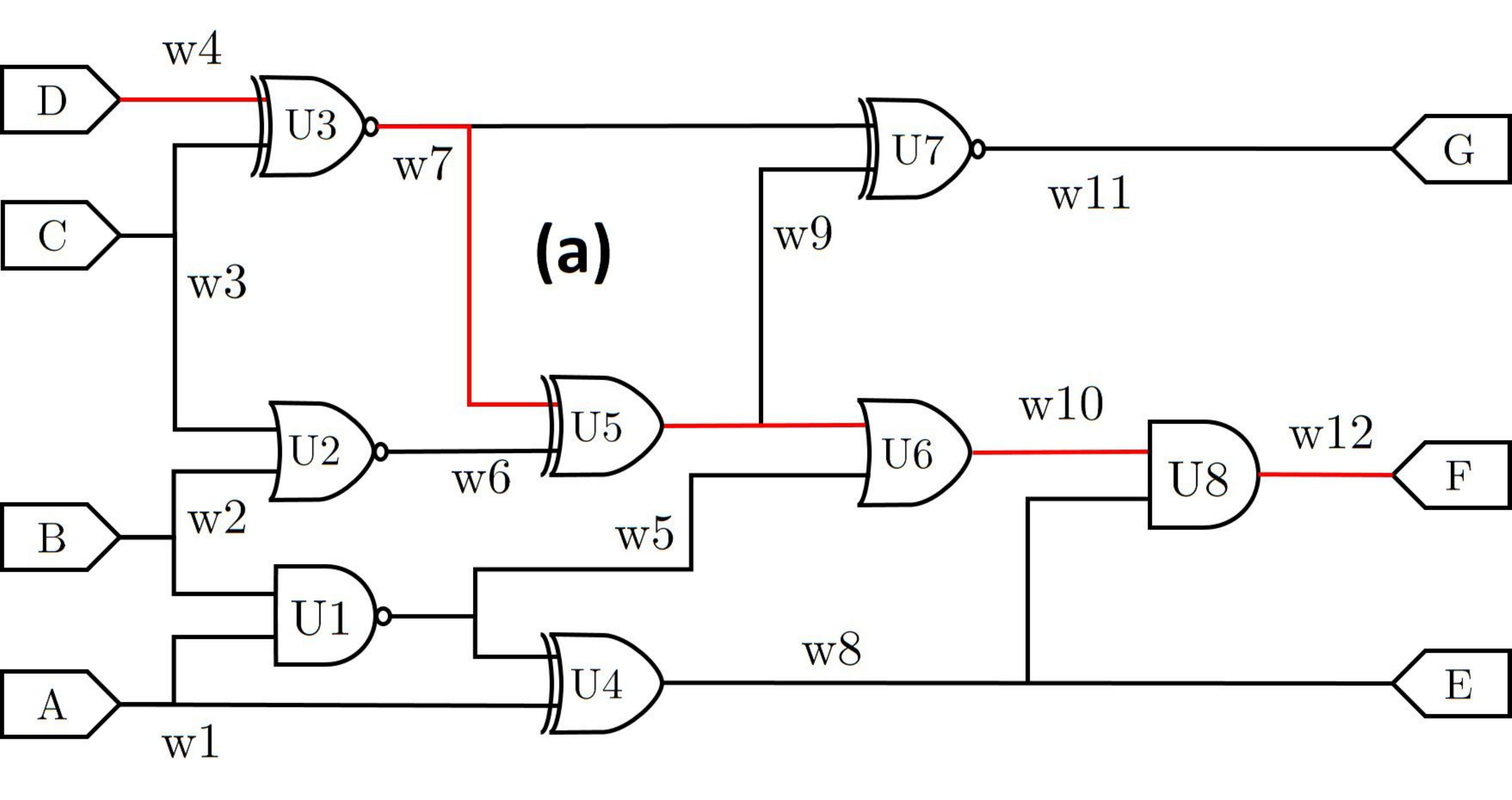}
    \label{fig:approx_acc}
    }
    \\
    \subfloat{
    \includegraphics[height=.45\textwidth, angle=90]{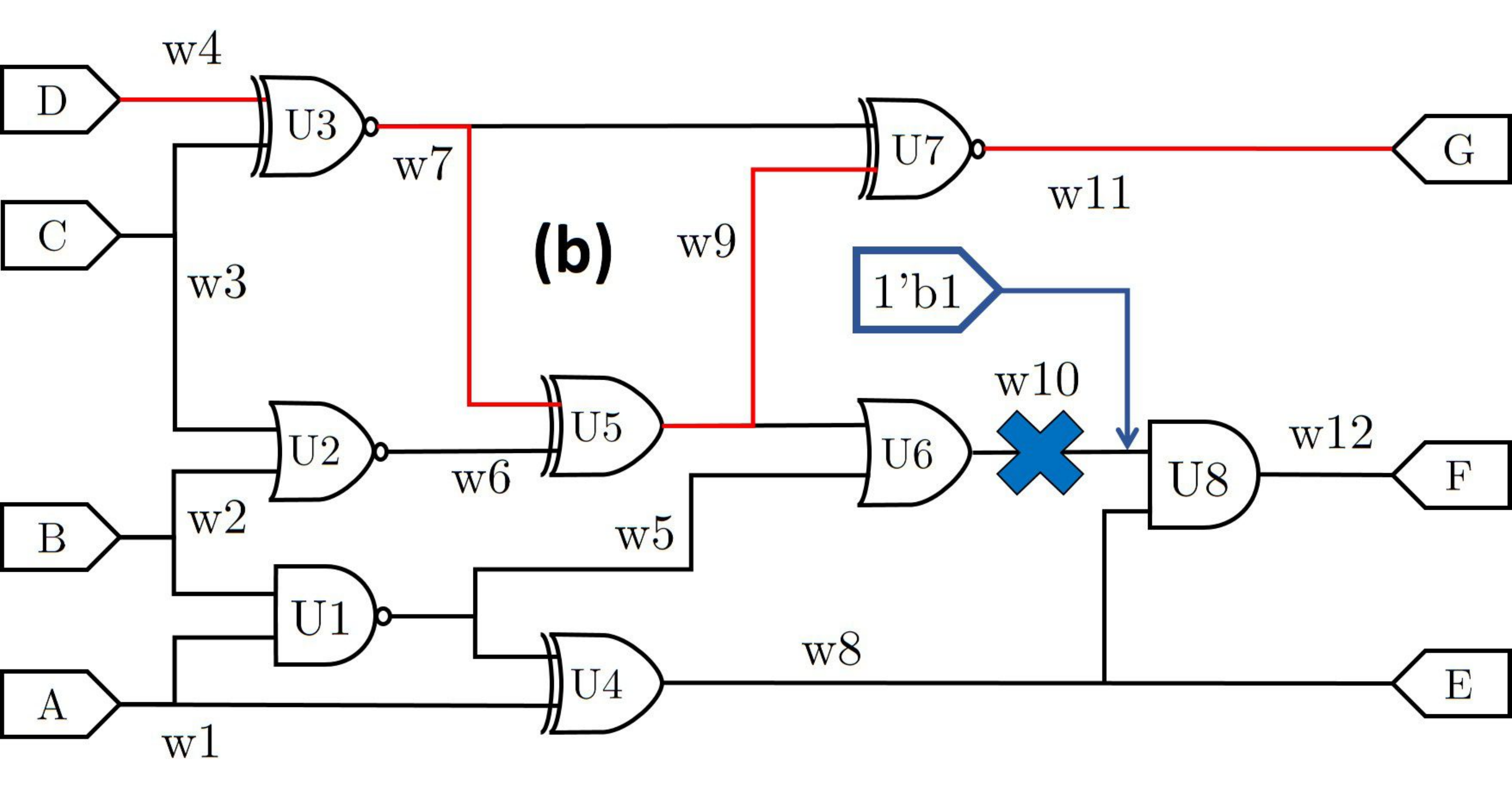}
    \label{fig:approx_const}
    }
    \\
    \subfloat{
    \includegraphics[height=.45\textwidth, angle=90]{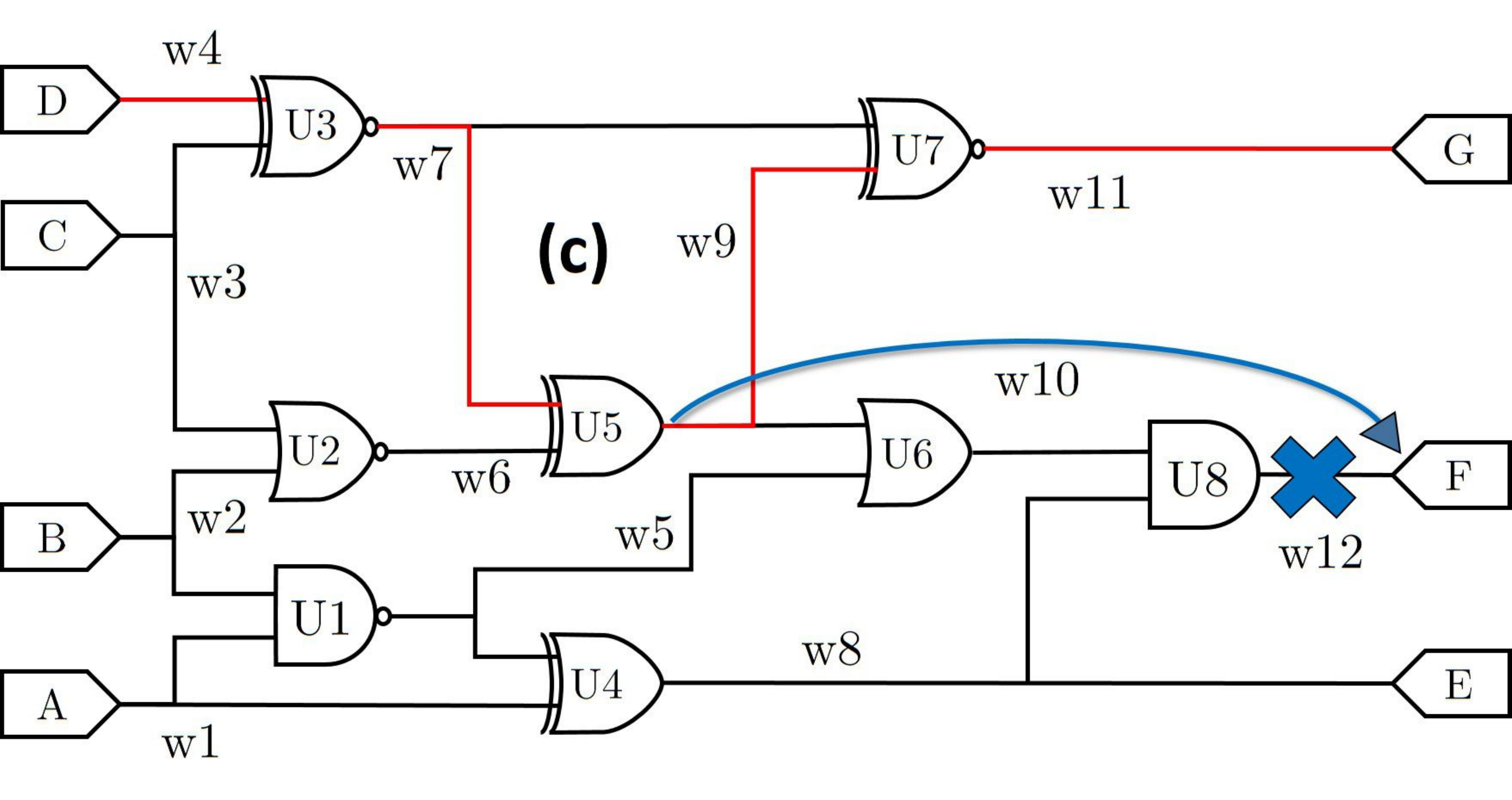}
    \label{fig:approx_wire}
    }
    \caption{Descriptive example of the wire-by-constant and wire-by-wire approximations employed in our framework.
    (a) The accurate netlist (DAG), (b) the approximate netlist (DAG) after applying wire-by-constant replacement, and (c) the approximate netlist (DAG) after applying wire-by-wire replacement.
    Edges marked in red correspond to the critical path of the circuit.}
    \label{fig:approx}
\end{figure}

\subsection{Netlist Approximation Library}\label{sec:netap}
In our analysis, the approximation library employed by our framework comprises two state-of-the-art techniques: wire-by-wire~\cite{Venkataramani:DATE:2013} and wire-by-constant~\cite{Schlachter:TransVLSI:2017} replacements.
Both techniques have been extensively used in design methodologies involving high level synthesis.
However, to the best of our knowledge our work is the first to employ them concurrently and in a coordinated manner, to enable delay-optimized netlist approximations.
Nevertheless, our framework is not bound to only these techniques and can be extended to support other gate-level netlist-specific approximation techniques.
Netlist approximations are facilitated by the fact that every combinational circuit can be represented by a Directed Acyclic Graph (DAG), in which the nodes are the netlist's gates and the edges are the wires that connect the gates.
For example, Fig.~\ref{fig:approx_acc} presents the DAG of a simple netlist, derived from a boolean expression.
The timing information of the aged circuit is carried over to the DAG nodes.
Hence, the critical path delay can be estimated as the maximum sum of gate delays among all paths in the DAG (annotated by red lines in Fig.~\ref{fig:approx_acc}).
Thus, netlist approximation techniques are easily applied, by just manipulating the DAG of the baseline netlist.

Wire-by-constant substitutes a wire in the netlist by a constant logic value.
In wire-by-constant replacement, the approximated wire (DAG edge) is disconnected from the output port of its parent gate (DAG node) and instead, it is driven by a constant `$1$' or `$0$'.
By replacing a wire with a constant, we break the signal propagation through that wire and thus, all the timing paths comprising that wire, become potentially faster.
Fig.~\ref{fig:approx_const} depicts an example of wire-by-constant approximation.
In this example, the replacement of wire $w10$ by the constant `$1$' is demonstrated, resulting in $13\%$ error rate.
All paths through the edge $U6 \longrightarrow U8$ are interrupted, allowing for an acceleration of the updated critical path (by $10.6\%$) through the rest of the nodes.

Wire-by-wire replaces a wire (namely approximated wire) in the netlist with another wire in the netlist (namely approximation wire).
In this replacement, the approximated wire (DAG edge) is disconnected from the output port of its parent gate (DAG node) and instead, it is connected to the respective output port of the parent gate (DAG node) of the approximation wire.
By applying wire-by-wire replacement, if the approximation wire is closer to the DAG root\footnote{The propagation delay to the approximation wire is smaller than the propagation delay to the approximated wire.}
than the approximated wire, then all the timing paths comprising the approximated wire potentially become faster.
Fig.~\ref{fig:approx_wire} depicts an example of wire-by-wire approximation, where $w9$ replaces $w12$. 
Paths which previously comprised the edge $U5 \longrightarrow U6$ are now significantly accelerated by omitting the propagation delay of gates $U6$ and $U8$.
Again, the DAG shows a CPD delay reduction of $10.6\%$.

\subsection{Approximation Candidates}\label{sec:apcandidates}

To apply wire-by-wire and wire-by-constant replacement, for each wire in the netlist, we need to identify a set of possible approximation candidates.
The extracted candidates will be next used in our optimization phase to generate the approximate netlist.

To obtain efficient approximation candidates, we use a circuit simulator and run a functional simulation of the given baseline netlist.
During the performed simulation, we capture the value of each wire in the netlist at every cycle.
Then, for each wire $i$, we calculate the percentage of cycles that it was `0' (namely  $T^i_0$) and the respective percentage that it was`1' (namely $T^i_1$)~\cite{Schlachter:TransVLSI:2017}.
In addition, for each wire $i$ and $j$, so that $j$ is closer to the DAG root than $i$, we calculate the similarity $S^i_j$ between the wires $i$ and $j$.
By similarity, we define the percentage of cycles in which $i$ and $j$ feature the same value.
For each wire $i$, we select as approximation candidate ($AX_i$) the value $0$, $1$, or $j$, $\forall j$, that introduces the smallest error, i.e., 
features the highest score $T^i_0$, $T^i_1$, or  $S^i_j$, $\forall j$:
\begin{equation}\label{eq:cand}
\begin{split}
AX_i= &\alpha: (\alpha,\gamma) \in \{(k,T^i_k), \forall k\} \cup \{(j,S^i_j), \forall j\}, \\
      &\gamma=\max\big(\{T^i_0, T^i_1\} \cup \{S^i_j, \forall j\}\big)
\end{split}
\end{equation}
The largest the $\gamma$ value is in~\eqref{eq:cand}, the smallest the introduced error rate at wire $i$ becomes.
For example, $T^i_0 = 0.95$ means that for 95\% of the time the wire $i$ was $0$.
Hence, by replacing $i$ with $0$, we obtain accurate results for the wire $i$ at 95\% of the time.
Similarly, if $S^i_j = 0.95$, then the wires $i$ and $j$ have the same value for 95\% of the time and replacing $i$ with $j$, results in $5$\% error rate at $i$.
To identify the approximation candidates we select the replacement with the highest similarity. 
In cases where both wire-by-constant and wire-by-wire replacement techniques produce equally adequate approximation candidates, i.e. significant similarity score, our algorithm prefers to apply wire-by-constant replacement.
By substituting a wire with a constant value, the signal propagation passing though that wire is completely cut off, leading thus to higher delay reduction than any wire-by-wire replacement.
In cases where more than one wire-by-wire replacement techniques produce significant similarity score, the wire that is closer to the DAG root (i.e., higher delay reduction is achieved) is selected.
In case of a tie, one of these wires is randomly selected.

Considering only one approximation candidate for each wire (as in~\eqref{eq:cand}), efficiently limits the size of the design space defined by netlist approximation using wire-by-wire and wire-by-constant techniques.
Assuming $n$ wires in the baseline netlist, the size of the approximation space is $2^n$ (i.e., each wire can be either accurate or replaced by its approximation candidate).
Alternatively, for each wire $i$, we can select a set of approximation candidates:
\begin{equation}\label{eq:candset}
AXS_i=\{k\,|\,T^i_k > C,\,k\in[0,1] \}\,\cup\,\{j\,|\,S^i_j>C, \forall j\},
\end{equation}
with $C$ a user defined constant (e.g., $C=0.80$).
This approach would give more flexibility to our optimization algorithm and enable more fine-grained approximation.
However, assuming $|AXS_i|>1$, using~\eqref{eq:candset} would explode the design space. 

\subsection{DAG}\label{sec:dag}
Our framework uses a DAG to represent the input netlist (i.e., input circuit) and perform the approximations.
We use Python to parse the Verilog description of the post-synthesis gate-level netlist and generate the circuit's DAG.
Each gate in the netlist is represented by a DAG node and each wire by a DAG edge.
Using the timing information obtained by the aging-aware STA, we annotate the DAG with the respective delays.
Although, aging degradation is workload dependent, when we run the STA, in order to ensure guardband elimination even under worst-case scenarios, we consider the highest degradation due to aging, i.e., $50$mV $V_{TH}$ degradation, for all the gates in the netlist (details in Section~\ref{sec:aginglibs}).
For each node in the DAG, we store its name (e.g., U1), its type (e.g., NAND), and its delay.
Those information are directly obtained from the Verilog description of the netlist and the STA.

As aforementioned, the employed approximation techniques are applied by directly manipulating the DAG.
After each approximation, the delay of the approximate netlist is obtained by traversing the respective approximate DAG~\cite{Zervakis:IEEEACC:2020}.
In addition, after each approximation, we translate the DAG to a C function\footnote{Our approximation library comprises only functional logic approximation techniques, enabling the DAG-to-C conversion.}.
To perform this conversion, we use the stored information for each node and we generate the C representation by hierarchically parsing the approximate DAG.
Each node is converted to the corresponding C function w.r.t the node type.
Since C code is serial, whereas Verilog is concurrent, the order in which the DAG nodes are written in the C function, is determined by the topological ordering of the nodes in the DAG.
The DAG edges (wires) are used as the inputs and outputs of the C function.
Leveraging this DAG-to-C conversion, we simulate the functionality of the approximate DAG in a very fast manner and we obtain a precise estimation of its error using large input datasets.
Note that, the goal of our framework is to completely eliminate the aging-induced timing guardbands.
As a result, for the final solution, the error of the fresh and aged approximate netlists will be the same and equal to the error calculated by the C-level error evaluation.

Alternatively, delay and error evaluation of the approximate DAG can be performed by converting back the DAG to Verilog and using the EDA tools to run STA and circuit simulation.
However, this would limit the scalability of our framework.
Commercial EDA tools are under license agreements and thus, the number of parallel evaluations is constrained by the number of available licenses.

\SetAlFnt{\footnotesize}
\begin{algorithm}[!t]

    \DecMargin{1pt}
    \SetFuncSty{cmss}
    \SetCommentSty{cmr}
    \SetKwComment{tcc}{\# }{}

    \SetKwInOut{Input}{Inputs}
    \SetKwInOut{Output}{Output}
    \SetKwFunction{init}{InitializePopulation}
    \SetKwFunction{select}{Selection}
    \SetKwFunction{cross}{Crossover}
    \SetKwFunction{mut}{Mutation}
    \SetKwFunction{fit}{CalcFitness}
    \SetKwBlock{Begin}{}{end}

    \Input{1. List of approximation candidates: \textit{C}, 
        2.  Delay target,
        3. DAG,
        4. Accurate netlist c-file: \textit{Netlist.c},
        5. Simulation inputs: \textit{Inputs.txt},
        6. GA parameters
    }
    \Output{Optimal approximate netlist}
    \BlankLine
    \tcc{Create random chromosomes, weighted by the critical path}
    Initialize population\;\label{algo:line:init}
    \For{ Generations }{\label{algo:line:for}
        \tcc{Produce new generation}
        Keep fittest solution\;\label{algo:line:elit}
        \While{NewGeneration incomplete}{\label{algo:line:while}
            Select parents\;
            Produce offsprings\;
            Add to new generation\;
        }\label{algo:line:endwhile}
        \tcc{Select the \textit{PopulationSize} fittest chromosomes}
        Calculate offspring fitness\;\label{algo:line:keeps}
        Merge parent and offspring populations\;
        Keep \textit{PopulationSize} fittest chromosomes\;\label{algo:line:keepf}
        Update MutationProbability if necessary\;\label{algo:line:mp}
    }
    \Return Population[0]\;\label{algo:line:return}
    \BlankLine
    \BlankLine
    \KwSty{Function} \fit{Chromosome, DelayTarget C, DAG, Netlist.c Inputs.txt}: \Begin{\label{algo:line:fitfunc}
    
        Decode: $Chr \longleftrightarrow C$ \;\label{algo:line:dec}
        Apply approximation on DAG\;\label{algo:line:approx}
        Isolate nodes with constant output\;\label{algo:line:isol}
        Update critical path\;\label{algo:line:cpd}
        \If{Delay $>$ DelayTarget}{\label{algo:line:target}
            \Return 0\;
        }\label{algo:line:targetend}
        
        Create approximate netlist\;\label{algo:line:cnetl}
        Calculate error\;\label{algo:line:calcerror}
        \Return $1/Error$\;\label{algo:line:returnerr}
    }\label{algo:line:endfitfunc}
    \caption{Heuristic Optimization Pseudocode}
    \label{algo:ga}
\end{algorithm}

\subsection{Genetic Optimization Algorithm}\label{sec:heur}
Our heuristic optimization procedure is described in Algorithm~\ref{algo:ga}.
Algorithm~\ref{algo:ga} implements a Genetic Algorithm to extract a good enough solution for~\eqref{eq:optimization}.
As inputs, Algorithm~\ref{algo:ga} receives i) the annotated DAG, ii) the list of approximation candidates, iii) the delay target (i.e., CPD of the baseline fresh netlist), and iv) the configuration parameters for the Genetic Algorithm.
For the representation of each chromosome in the population, we use bitstrings of length equal to the number of wires in the netlist.
Each bit indicates the use ($1$) or not ($0$) of the approximation candidate in the same position.
Bitstrings also help accelerate the Genetic Algorithm at runtime, since only two options are possible, i.e., each wire is either accurate or approximated.
The population is initialized with semi-random chromosomes that include approximation candidates for the gates belonging in the critical path (line \ref{algo:line:init}).
This enables Algorithm~\ref{algo:ga} to explore solutions of diminished delay from the first generation.
Throughout the standard iterative process, the Genetic Algorithm selectively produces offspring chromosomes until the next generation is populated (lines \ref{algo:line:while}-\ref{algo:line:endwhile}).
Leveraging the inherent parallel nature of the genetic algorithm, this process can be run on multiple threads concurrently, thus efficiently utilizing the underlying hardware to speedup the optimization phase.
An elitist approach is implemented (line \ref{algo:line:elit}), as the parent chromosomes of highest fitness are propagated intact to the next generation.
Depending on the new population's diversity, the mutation probability can be adaptively tuned to increase exploration (line \ref{algo:line:mp}).

The core of our optimization algorithm is the fitness calculation of approximate solutions (\ref{algo:line:fitfunc}-\ref{algo:line:endfitfunc}).
For any solution, the respective bitstring is decoded into wire-by-wire and wire-by-constant approximations (line \ref{algo:line:dec}), which are then used to modify (approximate) the baseline DAG (line \ref{algo:line:approx}).
Next, the error and delay of the approximate DAG are calculated (see Section~\ref{sec:dag}).
The obtained delay is compared with the delay target (lines \ref{algo:line:target}-\ref{algo:line:targetend}).
If the delay target is not met, the solution is discarded and considered invalid.
Only valid solutions are rewarded with a non-zero fitness score.
The score aims to promote approximations with minimum error, that satisfy the delay target.
To that end, the score is equal to the inverse of the error (line \ref{algo:line:returnerr}), since we treat the evolution as a maximization problem.
As an error metric for quantifying the induced approximation error, we use the Normalized Mean Error Distance (NMED)~\cite{Zervakis:TVLSI:2016}:
\begin{align}
NMED = \frac{1}{max}\frac{1}{N}\sum_{i=0}^N \frac{|Y_{true}-Y_{appr}|}{Y_{true}}
\label{eq:med}
\end{align}
where $Y_{true}$ is the accurate output, $Y_{appr}$ the approximate output, $N$ the number of inputs and $max$ is the maximum possible value.

The output of the Genetic algorithm is the solution with the highest fitness score, and therefore lowest possible error. 
Note that only solutions that meet the delay constraint are considered.
The obtained solution is used to approximate the baseline DAG.
Finally, the latter is synthesized using the EDA tool to exploit all the supported optimizations and remove any floating gates.
Note that, our framework targets combinational circuits.
Sequential circuits are out of the scope of this work.
Nevertheless, sequential circuits can be seamlessly supported by using the EDA tools to perform the error and delay evaluations in Algorithm~\ref{algo:ga}.

\section{Results and Evaluation}\label{sec:exp}
In this section, we evaluate the efficiency of our framework in eliminating the timing errors due to aging and thus to completely remove the aging-induced timing guardbands, boosting the performance of circuits.
Our framework is evaluated over varying arithmetic circuits as well as more complex dataflows.
Arithmetic circuits such as adders and multipliers are the basic building block of DSP and machine learning circuits~\cite{Zervakis:TVLSI:2019:vader}.
In our analysis, we consider an 8-bit and a 16-bit speed optimized adder from the industry-strength Synopsys DesignWare library~\cite{synopsys}.
Similarly, an 8-bit and a 16-bit speed optimized multiplier from DesignWare are examined.
Moreover, our framework is evaluated over five image processing benchmarks: Sobel Edge Detector~\cite{Zervakis:TVLSI:2019:vader}, 8-tap FIR filter~\cite{Paim:TCASI:2021}, Sum of Absolute Differences (SAD)~\cite{Paim:TCASI:2021}, Gaussian Blur filter (Blur)~\cite{Lee:Springer2019:ahls}, and a generic $3\times3$ Convolution kernel (Conv)~\cite{Lee:Springer2019:ahls}.
Targeting maximum performance, all the examined circuits (arithmetic units and image processing circuits) are synthesized with zero-slack.
Synopsys Design Compiler~\cite{synopsys} is used for circuit synthesis and the  \texttt{compile\_ultra} command is specified to obtain fully optimized netlists.
Synopsys PrimeTime~\cite{synopsys} is used to perform static time analysis (STA).
Mentor QuestaSim is used to run post-synthesis timing simulations and obtain the circuit's outputs.
The output of the timing simulation is used to calculate the circuit's error in the case of approximated and/or aged circuits.
For each circuit examined, two randomly generated input datasets are utilized.
The first consists of $10^5$ inputs and is used in the optimization phase of our framework.
The second one comprises $10^6$ inputs and it is used for the performed accuracy evaluation.
Input stimuli of both datasets are generated by randomly sampling a uniform distribution in the segment defined by the input bit-width of each circuit.

\begin{figure}[t]
    \centering
    \includegraphics[width=.48\textwidth]{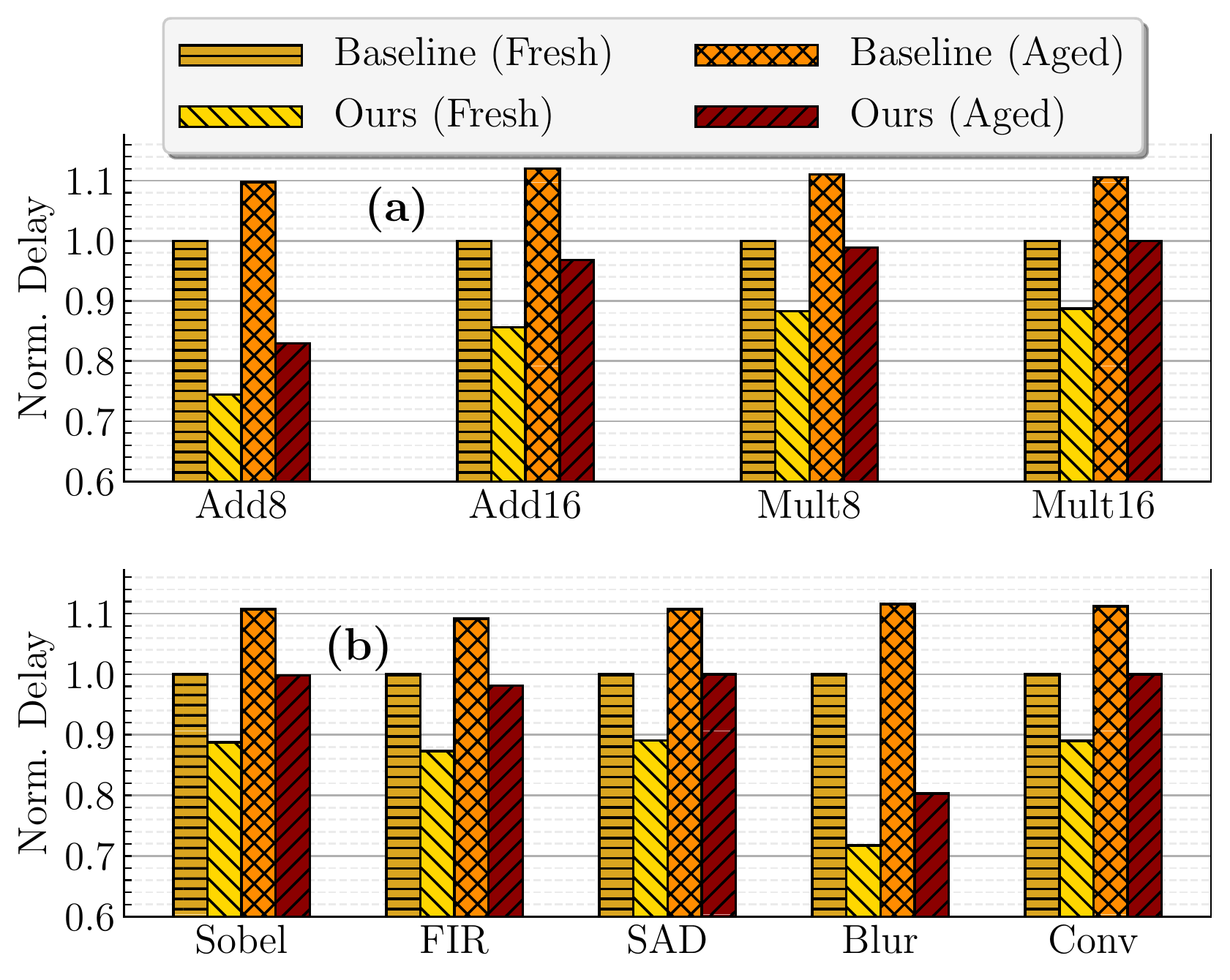}
    \caption{Delay evaluation for (a) the arithmetic circuits and (b) the image processing benchmarks. $50$mV $\Delta V_{th}$ degradation is considered for the aged circuits.
    As shown, the delay of our aged circuits is always less than or equal to the delay of the respective baseline fresh circuit. All the delay values reported are normalized w.r.t. the delay of the corresponding baseline fresh circuit.
    }
    \label{fig:delay}
\end{figure}

\subsection{Timing Guardbands Elimination}\label{sec:guardb}
First, we analyze the efficacy of our framework in eliminating the aging-induced timing guardbands.
Fig.~\ref{fig:delay} presents the delay evaluation of the baseline circuits and the approximate ones generated by our framework.
For each case, the delay of both the fresh as well as the aged ($\Delta V_{th}$=$50$mV) circuit is examined. 
All the delay values, reported in Fig.~\ref{fig:delay}, are normalized w.r.t. the delay of the respective baseline fresh circuit.
As shown in Fig.~\ref{fig:delay}, for all the examined circuits, the delay of the aged approximate circuit (generated by our framework) is less or equal to the delay of the corresponding baseline fresh circuit.
Hence, from day zero to the end of the projected lifetime (about $10$ years or $\Delta V_{th}$=$50$mV), our approximate circuits can be operated at the maximum frequency of the respective baseline fresh circuit with error guarantees, i.e., they will not exhibit timing errors due to aging.
As a result, by applying directed, aging-aware approximation, our framework enables operation at the maximum frequency for both the arithmetic circuits (Fig.~\ref{fig:delay}a) as well as the more complex image processing benchmarks (Fig.~\ref{fig:delay}b), eliminating thus the aging-induced timing guardbands of the baseline circuit.

\begin{figure}[t]
    \centering
    \includegraphics[width=.48\textwidth]{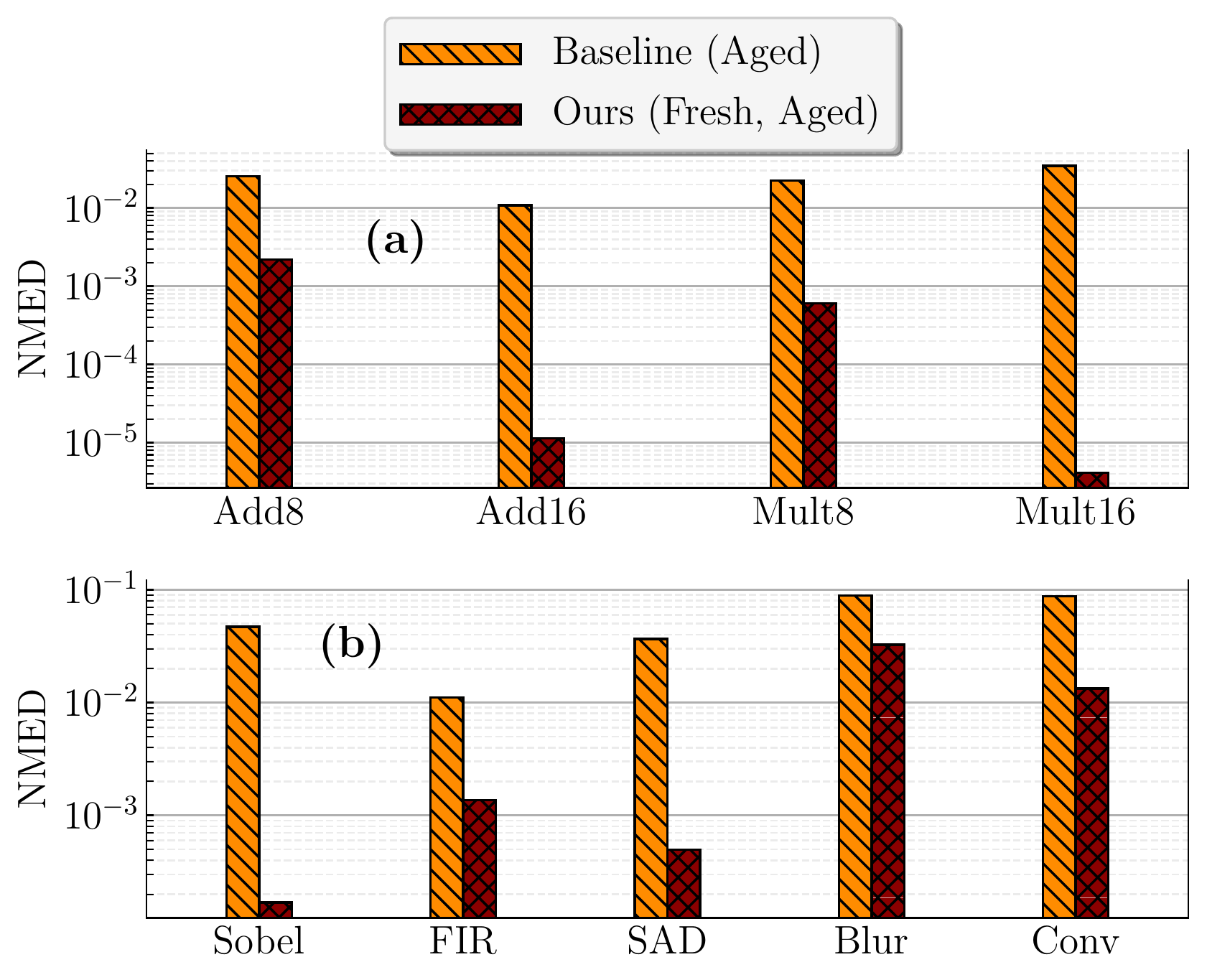}
    \caption{
    Iso-performance accuracy evaluation for (a) the arithmetic circuits and (b) the image processing benchmarks.
    No timing guardbands are considered and all the circuits are operated at their maximum frequency of the respective fresh baseline circuit.
    Our circuits do not feature aging-induced timing errors and thus their NMED is constant from the beginning to the end of the projected lifetime (i.e., $\Delta V_{th}$=$50$mV).
    }
    \label{fig:error}
\end{figure}

As demonstrated in Fig.~\ref{fig:delay}, the delay of the aged baseline circuits is on average $12.15$\% higher than the delay of the respective fresh baseline circuits.
Therefore, either a timing guardband will be used that will degrade the performance of the examined circuits by $12.15$\% on average, or aging-induced timing errors will occur.
In Fig.~\ref{fig:error} we evaluate the error value (NMED) of the baseline circuits when operated at maximum frequency, i.e., without any timing guardbands.
In addition, Fig.~\ref{fig:error} presents the error of our approximate circuits.
Note that, since the approximate circuits generated by our framework can operate at the maximum frequency of the baseline fresh circuit without any timing errors, their error value is constant for their entire projected lifetime (i.e., fresh and aged feature the same error).
As shown in Fig.~\ref{fig:error}, the approximate circuits generated by our framework achieve significantly higher accuracy than the aged baseline circuits.
For the case of arithmetic circuits (Fig.~\ref{fig:error}a), our approximate ones achieve on average $2340$x lower NMED than the respective aged baseline circuits.
As shown in Fig.~\ref{fig:error}a, the baseline circuits are affected by aging in a quite similar manner, i.e., they feature similar NMED values.
This is explained by the fact that timing errors (due to aging) appear in the slowest paths that mainly affect the most significant bits (MSBs) in arithmetic circuits~\cite{Salamin:DATE:2021}.
This is not the case for the approximate circuits, generated by our framework.
For the 16-bit circuits, NMED is significantly smaller compared to the 8-bit ones.
This is explained by the fact that the examined 8-bit circuits are quite small, featuring only a small number of wires.
Hence, approximating a wire in the 8-bit arithmetic circuits has higher impact on the final output compared to approximating a wire in the 16-bit arithmetic circuits.
Nevertheless, our framework achieves significantly lower error than the baseline for both the 8-bit and 16-bit circuits.
This error gain is also retained in the case of the examined image processing circuits (Fig.~\ref{fig:error}b).
In Fig.~\ref{fig:error}b, our approximate circuits achieve on average $74$x lower NMED than the respective aged baseline ones.
The NMED reduction is $280$x, $8$x, $74$x, $3$x, and $7$x for the Sobel, FIR, SAD, Blur and Conv benchmarks, respectively.
It is noteworthy that, in Fig.~\ref{fig:error}, for all the approximate circuits generated by our framework, the maximum NMED is $3.25\times10^{-2}$ and the average NMED is $5\times10^{-3}$.
Hence, for a negligible error of $5\times10^{-3}$ our framework is able to completely eliminate the aging-induced timing guardbands.
In addition, note that our framework induces a functional error, known at design time, that is constant from the beginning to the end of the projected lifetime.
On the other hand, errors due to aging are timing errors and depend on the input sequence and the state of the system, being  thus practically infeasible to predict.

\begin{figure}[t]
    \centering
    \includegraphics[width=.48\textwidth]{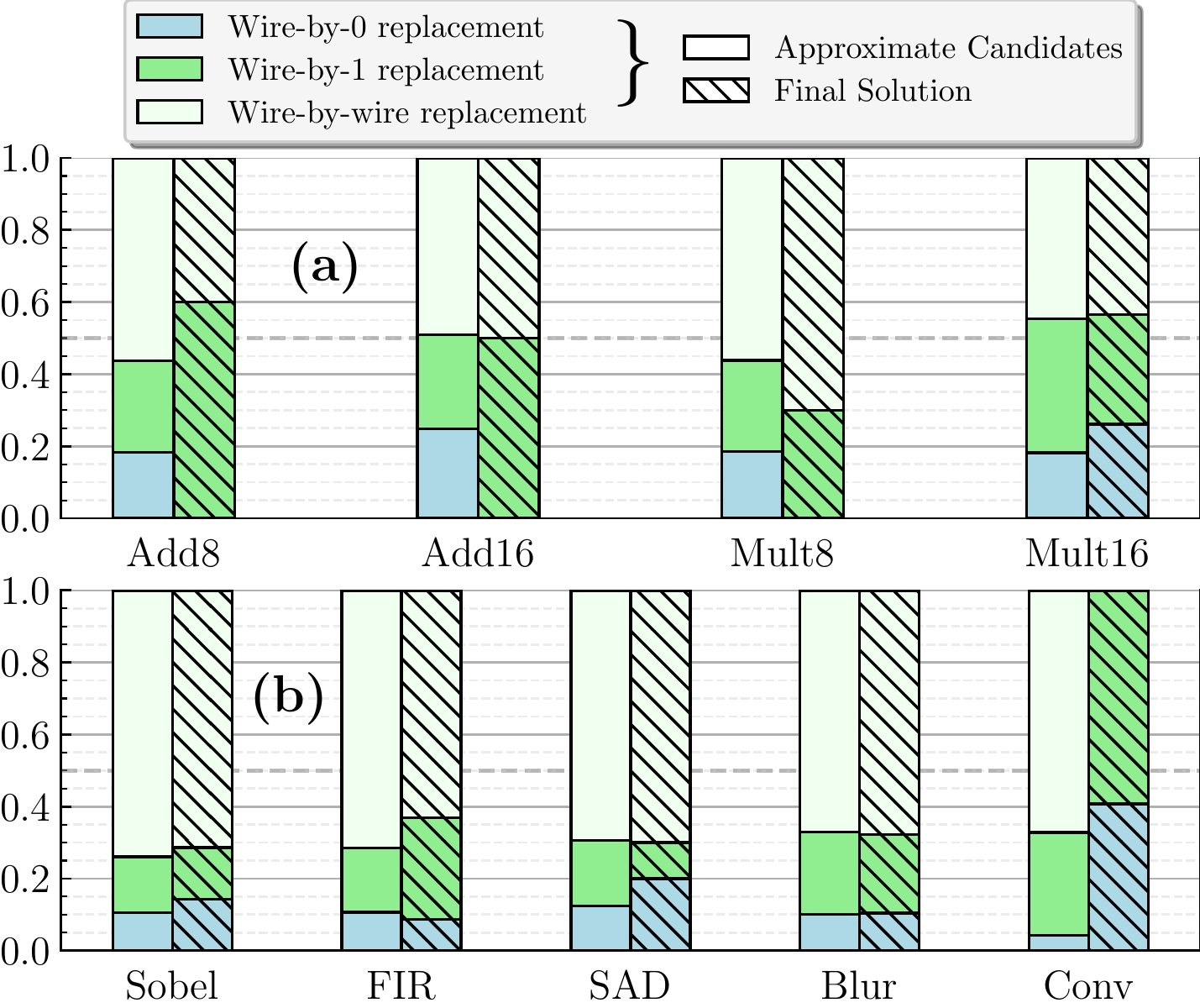}
    \caption{Percentage of selected approximation techniques w.r.t. the approximation candidates as well as the final approximate solution for (a) the arithmetic circuits and (b) the image processing benchmarks.
    Patterns distinguish between approximation candidates and final approximate solution, while bar colors indicate the approximation technique.}
    \label{fig:cand}
\end{figure}

Finally, we analyze our approximate circuits with respect to the applied approximation technique.
As mentioned in Section \ref{sec:apcandidates}, our framework considers both wire-by-wire and wire-by-constant approximation to generate aging-aware approximate circuits.
For each wire in the baseline netlist, an approximation candidate is selected that replaces the wire by either `0' or `1' (wire-by-constant) or by another wire in the netlist (wire-by-wire).
For each circuit examined, for the respective set of approximation cadidates,  Fig.~\ref{fig:cand} presents the percentage of candidates that employ wire-by-0, wire-by-1, and wire-by-wire replacement.
Similarly, for the final approximate netlist, Fig.~\ref{fig:cand} presents the percentage of selected approximations that employ wire-by-0, wire-by-1, and wire-by-wire replacement.
On average, $38.3\%$ of the approximation candidates employ wire-by-constant replacement while the remaining $61.7\%$ employ wire-by-wire replacement.
Specifically, for the case of arithmetic circuits, $48.5\%$ on average of the approximation candidates apply wire-by-constant replacement, while this value goes down to $30.2\%$ for the image processing circuits.
Considering the final approximate circuits generated by our framework, the $47.1\%$ of the applied approximations employ wire-by-constant replacement and the remaining $52.9\%$ employ wire-by-wire replacement.
Concluding, this almost balanced selection of wire-by-wire and wire-by-constant replacements, as well as the high diversity of the wire-by-0, wire-by-1, and wire-by-wire replacement percentages (as demonstrated in Fig.~\ref{fig:cand}) highlight the importance of considering both wire-by-wire and wire-by-constant techniques to build our final approximate circuit.

\begin{table}[t]
\caption{Comparative error (NMED) evaluation of our framework against state-of-the-art works for arithmetic circuits}.
\centering
{\footnotesize
\begin{tabular}{c|c|c|c|c}
    \hline
    & \textbf{Add8} & \textbf{Add16} & \textbf{Mult8} & \textbf{Mult16} \\ 
    \hline
    \textbf{APS~\cite{Kim:TCASI:2020, Amrouch:DAC:2017}} & $2.9\stimes10^{\text{-}3}$ & $3.0\stimes10^{\text{-}4}$ & $3.9\stimes10^{\text{-}3}$ & $6.9\stimes10^{\text{-}5}$ \\
    \hline
    \textbf{GLP~\cite{Schlachter:TransVLSI:2017}} &  $3.8\stimes10^{\text{-}2}$ & $4.5\stimes10^{\text{-}1}$ & $1.0\stimes10^{\text{-}1}$ & $2.4\stimes10^{\text{-}2}$ \\
    \hline
    \textbf{Ours} & $2.2\stimes10^{\text{-}3}$ & $1.1\stimes10^{\text{-}5}$ & $6.0\stimes10^{\text{-}4}$ & $4.1\stimes10^{\text{-}6}$ \\
    \hline
\end{tabular}
}
\label{tab:arith}
\end{table}

\begin{table}[t]
\caption{Comparative error (NMED) evaluation of our framework against state-of-the-art works for image processing benchmarks}.
\centering
{\footnotesize
\setlength{\tabcolsep}{3pt}
\begin{tabular}{c|c|c|c|c|c}
    \hline
    & \textbf{Sobel} & \textbf{FIR} & \textbf{SAD} & \textbf{Blur} & \textbf{Conv} \\ 
    \hline
    \textbf{APS~\cite{Kim:TCASI:2020, Amrouch:DAC:2017}} & $2.5 \stimes 10^{\text{-}2}$ & $2.3 \stimes 10^{\text{-}2}$ & $1.8 \stimes 10^{\text{-}3}$ & $1.2 \stimes 10^{\text{-}1}$ & $2.3 \stimes 10^{\text{-}2}$ \\
    \hline
    \textbf{DSEwam~\cite{Castro:ICCAD:2020:DSEwam}} & $8.8 \stimes 10^{\text{-}1}$ & $1.4 \stimes 10^{\text{-}1}$ & \text{-} & $1.6\stimes10^{\text{-}1}$ & \text{-} \\
    \hline
    \textbf{Ours} & $1.7 \stimes 10^{\text{-}4}$ & $1.4 \stimes 10^{\text{-}3}$ & $5.0 \stimes 10^{\text{-}4}$ & $3.3 \stimes 10^{\text{-}2}$ & $1.3 \stimes 10^{\text{-}2}$ \\
    \hline
\end{tabular}
}
\label{tab:bm}
\end{table}

\subsection{Comparison with State Of The Art}\label{sec:comp_sota}
In this section we compare our framework against state-of-the-art works.
In~\cite{Kim:TCASI:2020, Amrouch:DAC:2017} aging-aware precision scaling (APS) is used to generate an approximate multiplier and adder respectively.
To compare against APS, we employ the methodology presented in ~\cite{Kim:TCASI:2020, Amrouch:DAC:2017} and perform an exhaustive design space exploration to identify the optimal APS solution that satisfies the delay constraint.
Hence, using our exhaustive exploration, we apply APS to all the examined circuits.
Moreover, we compare our work against GLP~\cite{Schlachter:TransVLSI:2017}.
GLP is applied to arithmetic circuits and employs a systematic gate-level pruning.
GLP uses wire-by-constant replacement and in each iteration approximates the wires that minimize the significance-activity product (SAP).
Finally, we compare our work against DSEwam~\cite{Castro:ICCAD:2020:DSEwam}.
DSEwam targets more complex dataflows and thus it is used with the examined image processing benchmarks.
DSEwam uses analytical models to replace the exact functional units in a dataflow with faster approximate ones from an approximation library.
For APS, GLP, and DSEwam we use the same delay constraint as in our work, i.e., the delay of the aged ($\Delta V_{th}$=$50$mV) approximate circuit is less than or equal to the delay of the fresh baseline.

Tables~\ref{tab:arith} and~\ref{tab:bm} present the comparison of our work against the state-of-the-art ones for the examined arithmetic and image processing circuits, respectively.
As shown in Tables~\ref{tab:arith} and~\ref{tab:bm} our framework achieves always the highest accuracy.
In the case of arithmetic circuits (Table~\ref{tab:arith}), APS employs an aging aware approximation, achieving $472$x lower NMED than the aging-unaware GLP.
By applying directed fine-grained aging-aware approximation, our framework significantly outperforms both APS and GLP.
The approximate circuits generated by our framework feature on average $12.6$x and $11000$x times lower NMED than the APS and GLP approximate circuits, respectively.
Similar results are obtained for the image processing circuits in Table~\ref{tab:bm}.
The aging-unaware DSEwam features significantly higher NMED than the approximate circuits obtained with APS.
It is noteworthy that DSEwam wasn't able to generate a solution for the SAD and Conv circuits.
As shown in Table~\ref{tab:bm}, for all the examined circuits, our framework achieves on average $34$x lower NMED than APS.
The NMED reduction ranges from $1.3$x up to $145$x.
Compared to DSEwam, for the Sobel, FIR and Blur circuits, our framework delivers $1756$x lower NMED on average.

\begin{table}[t]
\caption{Execution time of a single offspring evaluation for the examined arithmetic and image processing circuits}
\label{tab:exec}
\centering
{\footnotesize
\renewcommand{\arraystretch}{1.2}
\begin{tabular}{c|c|c|c|c}
    \hline
    \multicolumn{5}{c}{\textbf{Arithmetic Circuits}}\\
    \hline
    \textbf{Circuit} & \textbf{Add8} & \textbf{Add16} & \textbf{Mult8} & \textbf{Mult16} \\
    \hline
    \textbf{Time [s]} & $0.198$ & $0.265$ & $0.495$ & $1.387$ \\
    \hline
    \textbf{\#Nodes (Gates)} & $71$ & $157$ &  $685$ & $1876$ \\
    \hline
\end{tabular}
\smallskip

\begin{tabular}{c|c|c|c|c|c}
    \hline
    \multicolumn{6}{c}{\textbf{Image Processing Benchmarks}}\\
    \hline
    \textbf{Circuit} & \textbf{Sobel} & \textbf{FIR} & \textbf{SAD} & \textbf{Blur} & \textbf{Conv}\\
   \hline
    \textbf{Time [s]} & $0.901$ & $0.768$ & $1.576$ & $0.618$ & $1.247$ \\
    \hline
    \textbf{\#Nodes (Gates)} & $1067$ & $906$ & $1893$ & $624$ & $1436$ \\
    \hline
\end{tabular}
}
\end{table}

\subsection{Execution Time Evaluation}\label{sec:exec_time}

Our framework implements a Genetic algorithm to address the optimization problem of aging-driven approximation as defined in~\eqref{eq:optimization}.
The time complexity of a Genetic algorithm is mainly defined by the number of epochs, the population size, the time complexity of one solution (i.e., the time required to evaluate one offspring) as well as the number of available threads.
Note that the optimization phase of our framework (see Fig.~\ref{fig:flow} and Algorithm~\ref{algo:ga}) is implemented purely in Python and C.
Hence, our framework i) can fully leverage the inherent parallel nature of the genetic algorithm allowing the optimization process to run concurrently on multiple threads exploiting all the available computing resources and ii) is not constrained by license limitations in commercial EDA tools that would restrict the scalability of our approach.
In Table~\ref{tab:exec}, we report the execution time required to evaluate one offspring for the examined arithmetic circuits (up) and image processing benchmarks (down). 
An offspring evaluation comprises both computing its error value as well as estimating its delay.
All experiments are conducted on a desktop computer featuring an AMD Ryzen 7 2700X processor at 3.2GHz and 32GB of RAM.
The values reported in Table~\ref{tab:exec} refer to the average value over 50 trials.
Overall, the execution time of a single offspring evaluation is very low for all the circuits, being at maximum $1.6$s for the largest examined circuit (i.e., SAD).
Still, as observed, the execution time is proportional to the circuit complexity (they are linearly correlated with an $R^2$ value of $0.972$).
The low execution times reported in Table~\ref{tab:exec} along with the scalability of our approach demonstrate the time-efficiency of our framework.
The number of epochs and the population size can be set as required by the user to control the overall execution time of our framework.
Nevertheless, note that running the algorithm for a few epochs, even though it will accelerate the optimization phase, it may potentially lead to solutions of lower quality.
In our paper, we have empirically set the population size and the number of epochs with respect to the circuit size.

\begin{table}[t!]
\caption{Power evaluation of the examined circuits.}
\label{tab:power}
    \centering
    {\footnotesize
    
    \begin{tabular}{c|c|c|c|c}
        \hline
        \multirow{2}{*}{\textbf{Circuit}}
        & \multicolumn{2}{c|}{\textbf{Power Fresh [W]}} &
        \multicolumn{2}{c}{\textbf{Power Aged [W]}} 
        \\ \cline{2-5}
        & \textbf{Baseline} & \textbf{Ours} & \textbf{Baseline} & \textbf{Ours}
        \\ \hline
        \textbf{Add8} & $1.00\times10^{\text{-}3}$ &  $9.79\times10^{\text{-}4}$ & $9.82\times10^{\text{-}4}$ & $9.58\times10^{\text{-}4}$
        \\ \hline
        \textbf{Add16} & $1.49\times10^{\text{-}3}$ & $1.80\times10^{\text{-}3}$ & $1.46\times10^{\text{-}3}$ & $1.76\times10^{\text{-}3}$
        \\ \hline
        \textbf{Mult8} & $3.30\times10^{\text{-}3}$ & $3.21\times10^{\text{-}3}$ & $3.21\times10^{\text{-}3}$ & $3.12\times10^{\text{-}3}$
        \\ \hline
        \textbf{Mult16} & $1.07\times10^{\text{-}2}$ &  $1.38\times10^{\text{-}2}$ &  $1.03\times10^{\text{-}2}$ & $1.33\times10^{\text{-}2}$
        \\ \hline
        \textbf{Sobel} & $6.04\times10^{\text{-}3}$ & $7.97\times10^{\text{-}3}$ & $5.84\times10^{\text{-}3}$ & $7.72\times10^{\text{-}3}$
        \\ \hline
        \textbf{FIR} & $5.63\times10^{\text{-}3}$ & $3.89\times10^{\text{-}3}$ & $5.45\times10^{\text{-}3}$ & $3.77\times10^{\text{-}3}$
        \\ \hline
        \textbf{SAD} & $8.68\times10^{\text{-}3}$ & $7.84\times10^{\text{-}3}$ & $8.54\times10^{\text{-}3}$ & $7.60\times10^{\text{-}3}$
        \\ \hline
        \textbf{Blur} & $4.45\times10^{\text{-}3}$ &  $1.09\times10^{\text{-}3}$ & $4.43\times10^{\text{-}3}$ & $1.07\times10^{\text{-}3}$
        \\ \hline
        \textbf{Conv} & $7.05\times10^{\text{-}3}$ &  $5.87\times10^{\text{-}3}$ & $6.83\times10^{\text{-}3}$ & $5.69\times10^{\text{-}3}$
        \\ \hline
    \end{tabular}
    }
\end{table}

\subsection{Power Discussion}\label{sec:power}
As described in Section~\ref{sec:framew}, the aim of our framework is to inject the bare minimum approximation so that aging-induced timing errors are eliminated and the error imposed by the applied logic approximation is minimized.
Hence, although power reduction may be achieved due to the applied approximation, it is out of the scope of the respective optimization decision~\eqref{eq:optimization} of our work.
For the sake of completeness, this section presents a power analysis over the examined circuits.
Table~\ref{tab:power} presents the power consumption of the baseline and our approximate circuits under both fresh and aged conditions.
To measure the power consumption, for each circuit, the switching activity obtained from the gate-level timing simulation is fed to Synopsys PrimeTime to run the power analysis.
Note that, in Table~\ref{tab:power}, the power of the aged circuits is lower than the power of the respective fresh ones.
This is explained by the fact that aging increases the threshold voltage of underlying transistors and thus both leakage and dynamic power become less as it has been demonstrated in~\cite{Amrouch:IRPS2017}.
As shown in Table~\ref{tab:power}, the circuits generated by our framework feature mainly lower power consumption than the corresponding baseline circuits.
However, three out of the nine circuits generated by our framework (i.e., the 16-bit adder, the 16-bit multiplier, and the Sobel) exhibit higher power consumption than the baseline.
For these circuits our framework achieved the highest error reduction (as shown in Fig.~\ref{fig:error}) while still satisfying the delay constraint of~\eqref{eq:optimization}.
As a result of inducing minimal approximation but also significantly decreasing the CPD of the circuit, a higher power consumption is attained.
Hence, to satisfy both the delay constraint of~\eqref{eq:optimization} as well as that the power consumption decreases, we need to reformulate our optimization problem as follows:
\begin{equation}\label{eq:optimization_pow}
\begin{gathered}
\text{given } fresh\, BL\, netlist \text{ find } aged\,AX\,netlist \\
\text{s.t.}\,\mathrm{CPD}(aged\,AX\,netlist)\!\leq\!\mathrm{CPD}(fresh\, BL\, netlist),\\
\mathrm{Power}(fresh\,AX\,netlist)\!\leq\!\mathrm{Power}(fresh\, BL\, netlist),\\
\text{and}\,\min\big(Error(aged\,AX\,netlist)\big).
\end{gathered}
\end{equation}
Nevertheless, this is a different optimization problem and doesn't ensure minimum error due to logic approximation, i.e., our optimization goal in~\eqref{eq:optimization}.
To address~\eqref{eq:optimization_pow}, a multi-objective heuristic solver as in~\cite{Mrazek:DATE:2017} may be employed, along with our delay and error modeling, and a high-level power estimation similar to~\cite{Zervakis:IEEEACC:2020}.

In Table~\ref{tab:new_sols} we present some dominated solutions generated by our framework during the optimization phase for the 16-bit adder, 16-bit multiplier, and Sobel.
The approximate circuits presented in Table~\ref{tab:new_sols}, satisfy the delay constraint of our optimization problem~\eqref{eq:optimization} and also achieve lower power consumption than the respective baseline.
These offsprings were generated and evaluated by our Genetic algorithm but they were eventually discarded later since they constitute dominated solutions as a better solution (i.e., with lower error) was found in the next iterations.
Note that the error of the aged baseline is $1.08\times10^{-2}$, $3.47\times10^{-2}$ and $4.69\times10^{-2}$ for the 16-bit adder, 16-bit multiplier and Sobel, respectively.
Therefore, the circuits of Table~\ref{tab:new_sols} achieve lower power as well as $168$x, $810$x and $3$x lower error, respectively, than the corresponding baseline.
Nevertheless, the circuits of Table~\ref{tab:new_sols} feature $5$x, $10$x and $81$x higher error than the final non-dominated solutions generated by our framework (see Fig.~\ref{fig:delay}-\ref{fig:error} and Table~\ref{tab:power}).
Hence, although Tables~\ref{tab:power} and~\ref{tab:new_sols} demonstrate that there exists at least one solution for \eqref{eq:optimization_pow}, the circuits of Table~\ref{tab:new_sols} are not solutions of our optimization problem as defined by~\eqref{eq:optimization}.

\begin{table}[t!]
\caption{Dominated solutions generated by our framework for the 16-bit adder, 16-bit multiplier, and Sobel that feature both lower critical path delay as well as lower power than the corresponding baseline.}
    \label{tab:new_sols}
    \centering
    {\footnotesize
    
    \setlength{\tabcolsep}{4pt}
    \begin{tabular}{c|c|c|c|c|c}
        \hline
        \multirow{2}{*}{\textbf{Circuit}}
        & \multicolumn{2}{c|}{\textbf{Norm. Delay}} &
        \multirow{2}{*}{\textbf{Error (NMED)}} &
        \multicolumn{2}{c}{\textbf{Power [W]}}
        \\ \cline{2-3} \cline{5-6}
        & \textbf{Fresh} & \textbf{Aged} & &\textbf{Fresh} & \textbf{Aged}
        \\ \hline
        \textbf{Add16} & $0.888$ & $0.992$ & $6.44\times10^{-5}$ & $1.3\times10^{-3}$ & $1.32\times10^{-3}$
        \\ \hline
        \textbf{Mult16} & $0.871$ & $0.987$ & $4.28\times10^{-5}$ & $1.04\times10^{-2}$ & $1.00\times10^{-2}$
        \\ \hline
        \textbf{Sobel} & $0.886$ & $0.996$ & $1.38\times10^{-2}$ & $5.58\times10^{-3}$ & $5.39\times10^{-3}$
        \\ \hline
    \end{tabular}
    }
\end{table}

\subsection{Process Variation Discussion}\label{sec:montecarlo}
In this section we evaluate the impact that process variations might have on the approximation efficiency and accuracy of our framework.
To achieve this, we perform a Monte Carlo analysis with $1000$ samples to capture the impact of variation on the output error of the examined circuits. 
It is noteworthy that, because in our work we target complex industrial circuits, generated and optimized using mature EDA tool flows (i.e.,~Synopsys Design Compiler and DesignWare library), Monte Carlo analysis using SPICE simulations is infeasible due to the vast time that would be required.
To overcome this challenge, a Monte Carlo analysis is conducted at the circuit level employing the detailed delay profile of the circuit (i.e.,~SDF file) as extracted from Static Timing Analysis using mature tool flows (i.e.,~Synopsys PrimeTime).
Note that for all examined circuits, both the baseline aged and the approximate aged are analyzed under variability effects for fair comparisons. 
Each circuit is subjected to $1000$ different variations (i.e., $1000$ different SDF files are generated per circuit), by modifying the timing characteristics of its composing gates.
To generate a variation (modified) SDF file, we parse the original SDF file of the circuit as generated by PrimeTime using the aged library and for each gate in the netlist we replace its delay $\delta$ with a new value $\delta^\prime$.
The value of $\delta^\prime$ is obtained by randomly sampling a normal distribution $\mathcal{N}(\delta, 0.01\delta^2)$.
The considered normal distribution has ($\sigma/\mu = 10\%$), which has been demonstrated in~\cite{Amrouch:TCAS1:2020} to be sufficient to capture the effects of different variability sources such as the gate work-function, channel length, geometrical variation, and oxide thickness. 
Finally, for each circuit and each modified SDF file, we run a timing simulation using Mentor Questasim and capture the output error.
All circuits are simulated at the CPD of the respective fresh baseline at nominal process parameters.
Overall, to perform the Monte Carlo analysis for all the examined circuits, $18000$ gate-level timing simulations are conducted.
\begin{figure}[t]
    \centering
    \includegraphics[width=\columnwidth]{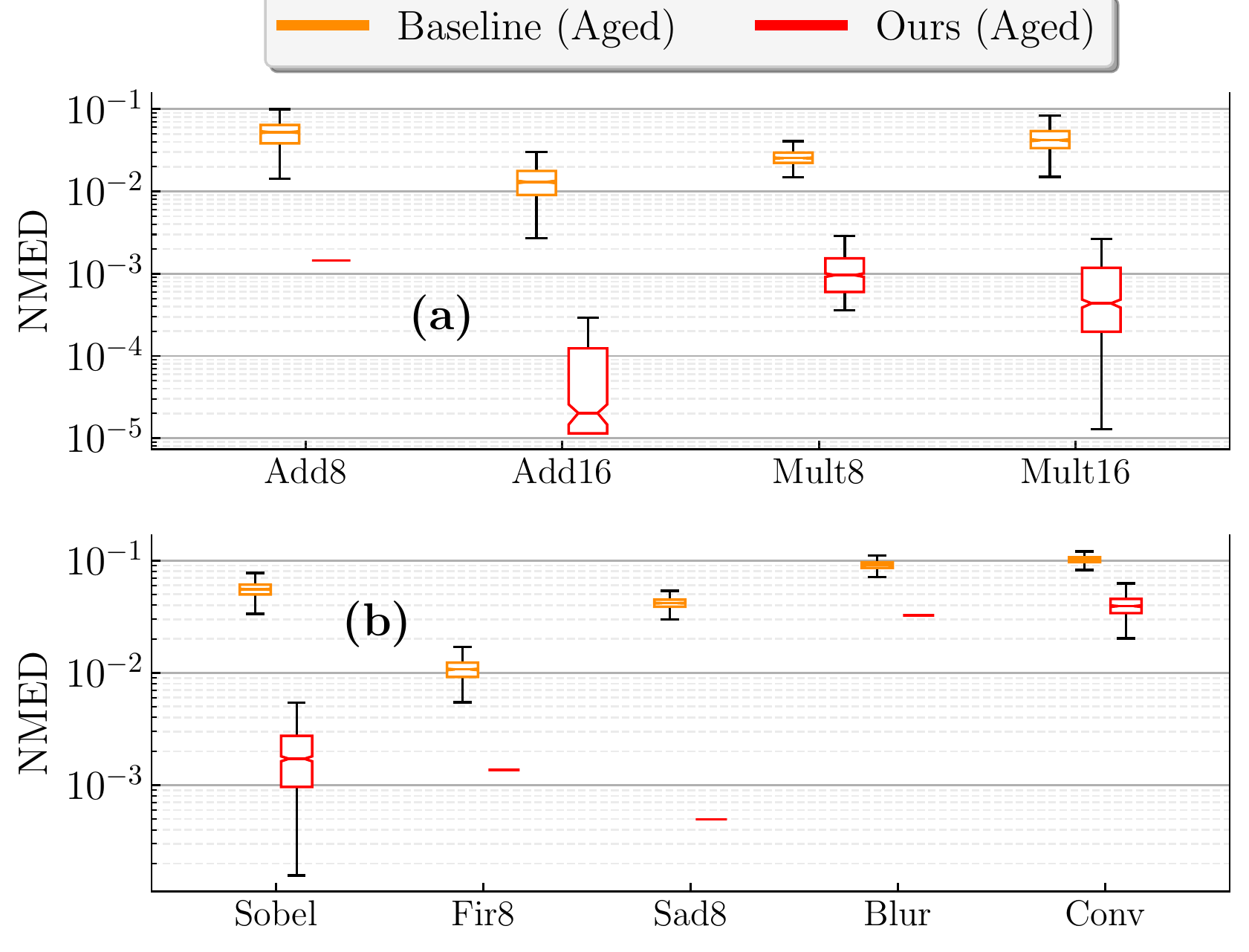}
    \caption{Output error from Monte Carlo error analysis with $1000$ samples of (a) arithmetic circuits and (b) image processing benchmarks.}
    \label{fig:mc_an}
\end{figure}

Fig.~\ref{fig:mc_an} summarizes the outcome of the Monte Carlo analysis.
For each circuit, the output error (NMED) is presented as a box plot.
Note that in the aged baseline circuits the output error is only due to the aging-induced timing errors, while in our aged approximate circuits the output error is subject to both timing errors and functional approximation.
As shown in Fig.~\ref{fig:mc_an}, the error of our aged circuit is always lower than the error of the baseline aged circuit.
Specifically, the IQ2 error of our approximate aged circuits is $40$x lower, on average, than that of the aged baseline.
Overall, the error reduction in our approximate aged circuits ranges from $1.31$x up to $6446$x.
In other words, despite process variation effects, by applying functional approximation in a directed manner, our framework is still able to significantly decrease the aging-induced error.
In addition, it is noteworthy that four out of the nine ($44$\%) circuits generated by our framework feature zero error variance and thus do not exhibit any timing errors.
Hence, by setting tighter delay constraints in~\eqref{eq:optimization} we can completely eliminate the timing errors for all the circuits examined. 
However, in the latter case our framework would have to apply higher approximation and thus induce a larger static functional error (compared to Fig.~\ref{fig:error}).
Though, studying and mitigating process variation effects is out of the scope of our paper.

\section{Conclusion}\label{sec:conc}
In this work, we propose and implement the first automated framework for aging-aware circuit approximation.
Our framework employs a genetic algorithm to apply wire-by-wire and wire-by-constant approximation in a systematic manner and suppress aging circuit degradation.
By inducing a small and known a-priori functional error, our framework manages to eliminate aging-induced timing errors and thus remove aging-induced timing guardbands, boosting the circuit's performance.
Finally, we demonstrate that, targeting to eliminate aging-induced timing guardbands, our framework significantly outperforms state-of-the-art approaches by delivering orders of magnitude lower error.

\begin{IEEEbiography}[{\includegraphics[width=1in,height=1.25in,clip,keepaspectratio]{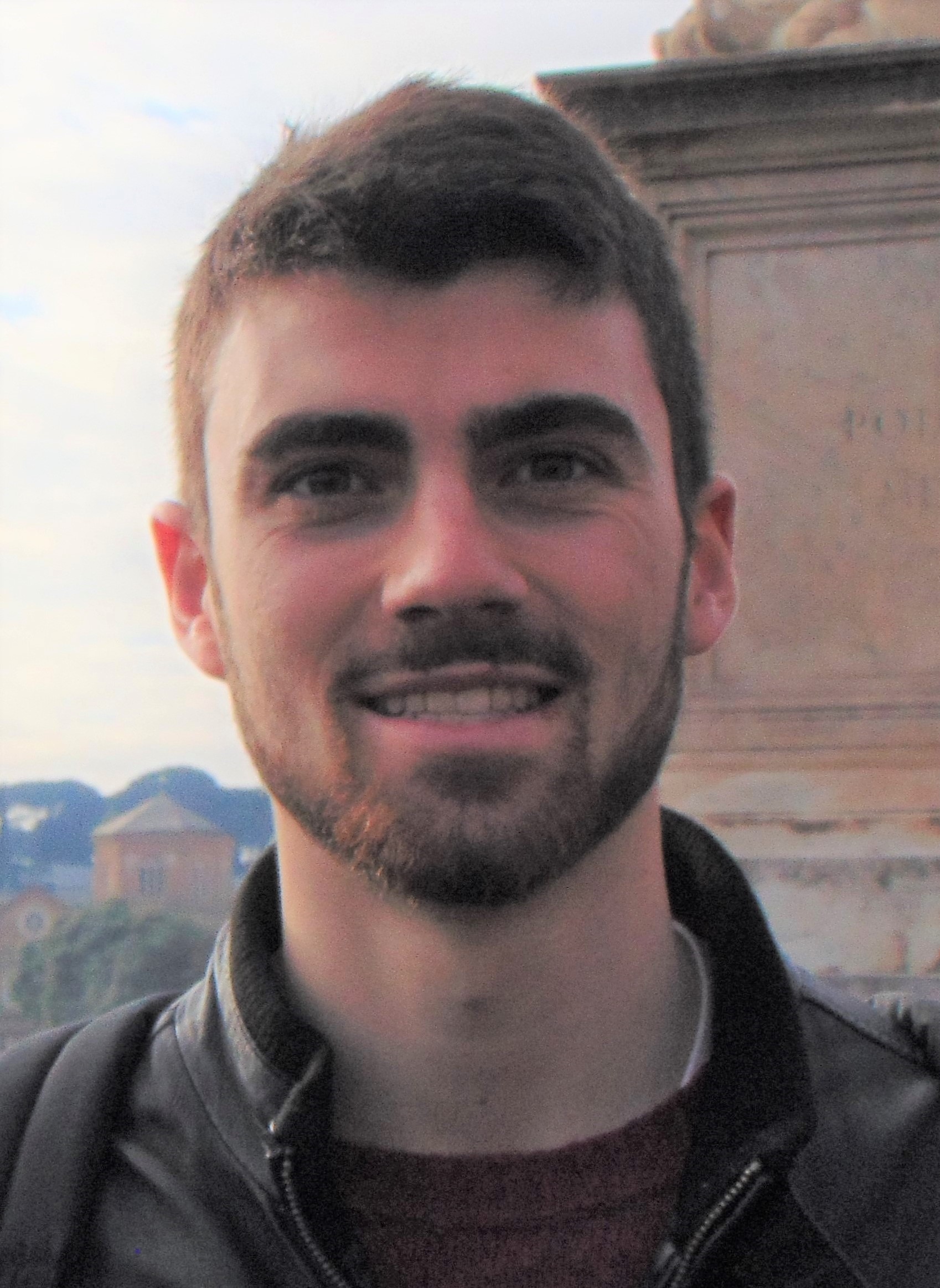}}]{Konstantinos Balaskas} received his Bachelor Degree in Physics and Master Degree in Electronic Physics from the Aristotle University of Thessaloniki in 2018 and 2020, respectively. Currently, he is a pursuing the PhD degree at the same institution. His research interests include approximate computing, machine learning and digital circuit design and optimization.
\end{IEEEbiography}

\begin{IEEEbiography}[{\includegraphics[width=1in,height=1.25in,clip,keepaspectratio]{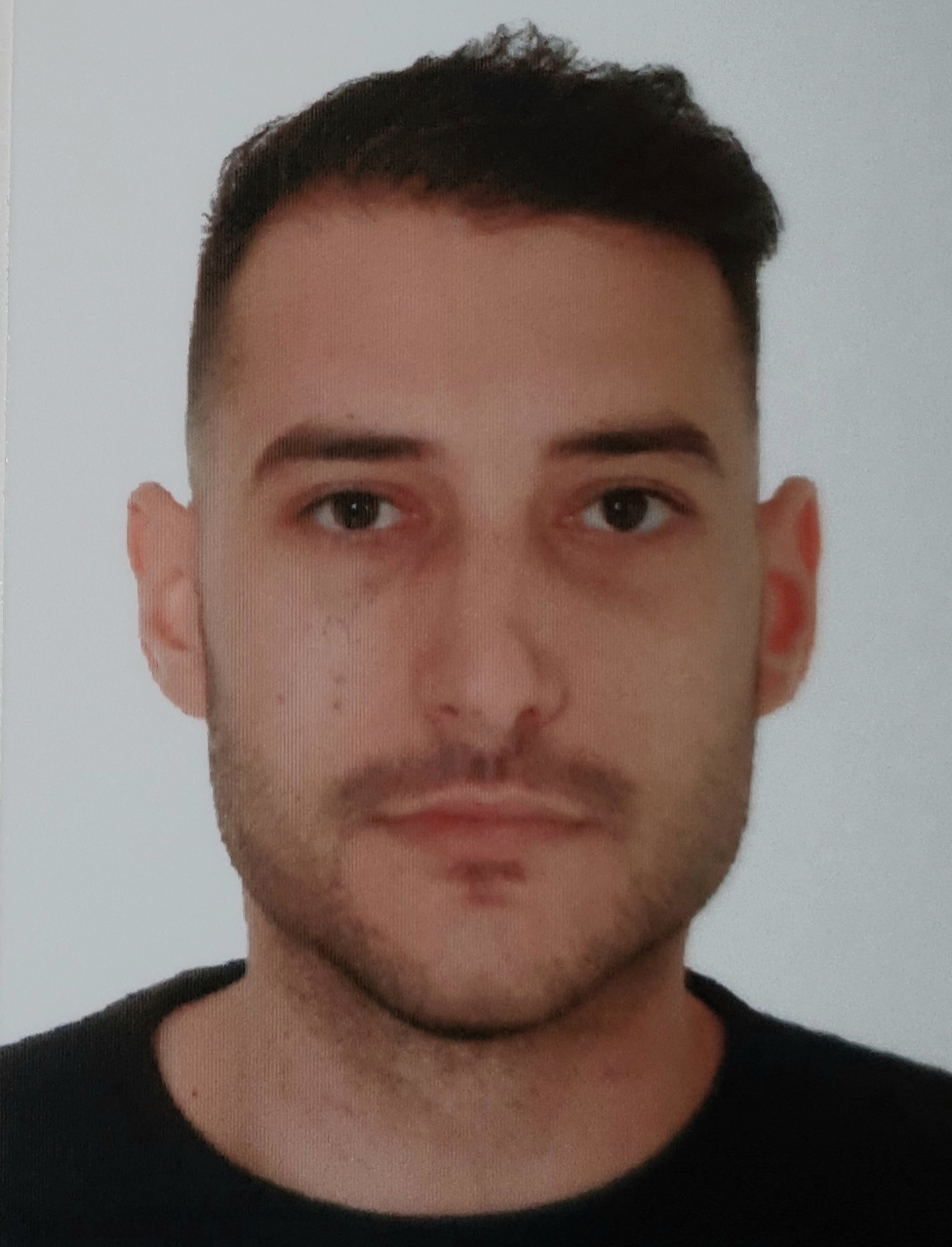}}] {Georgios Zervakis} is a Research Group Leader at the Chair for Embedded Systems (CES) at the Karlsruhe Institute of Technology (KIT), Germany.
He received the Diploma and the Ph.D. degree from the Department of Electrical and Computer Engineering (ECE), National Technical University of Athens (NTUA), Greece, in 2012 and 2018, respectively.
Before joining KIT, Georgios worked as a primary researcher in several EU-funded projects as member the Institute of Communication and Computer Systems (ICCS), Athens, Greece.
His research interests include approximate computing, low power design, design automation, and integration of hardware acceleration in cloud.
\end{IEEEbiography}

\begin{IEEEbiography}[{\includegraphics[width=1in,height=1.25in,clip,keepaspectratio]{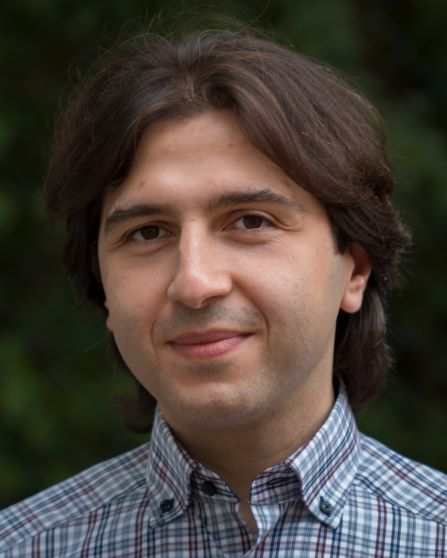}}] {Hussam Amrouch}(S'11-M'15) is a Junior Professor for the Semiconductor Test and Reliability (STAR) chair within the Computer Science, Electrical Engineering Faculty at the University of Stuttgart as well as a Research Group Leader at the Karlsruhe Institute of Technology (KIT), Germany. He received his Ph.D. degree with distinction (Summa cum laude) from KIT in 2015. His main research interests are design for reliability and testing from device physics to systems, machine learning, security, approximate computing, and emerging technologies with a special focus on ferroelectric devices. He holds seven HiPEAC Paper Awards and three best paper nominations at top EDA conferences: DAC'16, DAC'17 and DATE'17 for his work on reliability. He currently serves as Associate Editor at Integration, the VLSI Journal. He has served in the technical program committees of many major EDA conferences such as DAC, ASP-DAC, ICCAD, etc. and as a reviewer in many top journals like T-ED, TCAS-I, TVLSI, TCAD, TC, etc. He has 120 publications in multidisciplinary research areas across the entire computing stack, starting from semiconductor physics to circuit design all the way up to computer-aided design and computer architecture. He is a member of the IEEE. ORCID 0000-0002-5649-3102.
\end{IEEEbiography}

\begin{IEEEbiography}[{\includegraphics[width=1in,height=1.25in,clip,keepaspectratio]{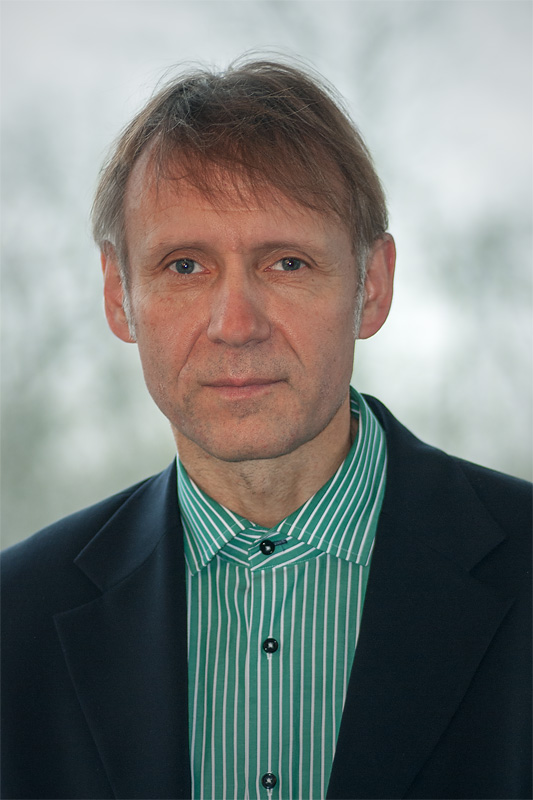}}]{J\"org Henkel} (M'95-SM'01-F'15)
is the Chair Professor for Embedded Systems at Karlsruhe Institute of Technology. Before that he was a research staff member at NEC Laboratories in Princeton, NJ. He received his diploma and Ph.D. (Summa cum laude) from the Technical University of Braunschweig. His research work is focused on co-design for embedded hardware/software systems with respect to power, thermal and reliability aspects. He has received six best paper awards throughout his career from, among others, ICCAD, ESWeek and DATE. For two consecutive terms he served as the Editor-in-Chief for the ACM Transactions on Embedded Computing Systems. He is currently the Editor-in-Chief of the IEEE Design\&Test Magazine and is/has been an Associate Editor for major ACM and IEEE Journals. He has led several conferences as a General Chair incl. ICCAD, ESWeek and serves as a Steering Committee chair/member for leading conferences and journals for embedded and cyber-physical systems. Prof. Henkel coordinates the DFG program SPP 1500 ``Dependable Embedded Systems'' and is a site coordinator of the DFG TR89 collaborative research center on ``Invasive Computing''. He is the chairman of the IEEE Computer Society, Germany Chapter, and a Fellow of the IEEE.
\end{IEEEbiography}

\begin{IEEEbiography}[{\includegraphics[width=1in,height=1.25in,clip,keepaspectratio]{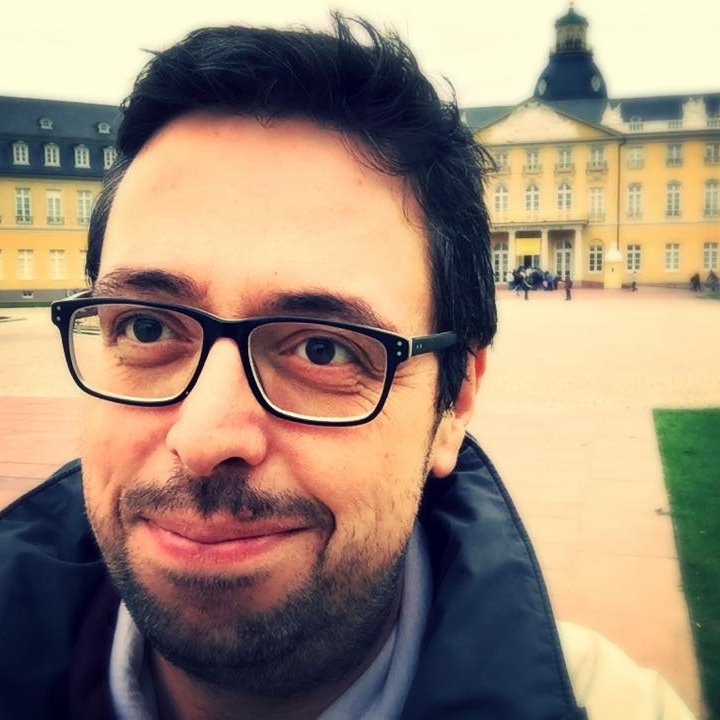}}]{Kostas Siozios} received his Diploma, Master and Ph.D. Degree in Electrical and Computer Engineering from the Democritus University of Thrace, Greece, in 2001, 2003 and 2009, respectively. Currently, he is Assistant Professor at Department of Physics, Aristotle University of Thessaloniki. His research interests include Low-Power Hardware Accelerators, Resource Allocation, Machine Learning, Decision-Making Algorithms, Cyber-Physical Systems (CPS) and IoT for Smart-Grid. He has published more than 130 papers in peer-reviewed journals and conferences. Also, he has contributed in 5 books of Kluwer and Springer. The last years he works as Project Coordinator, Technical Manager or Principal Investigator in numerous research projects funded from the European Commission (EC), European Space Agency (ESA), as well as the Greek Government and Industry.
\end{IEEEbiography}

\end{document}